\documentclass[10pt,twocolumn,twoside]{IEEEtran}

\usepackage{amsmath}
\usepackage{amssymb}
\usepackage{array}
\usepackage{fixmath}
\usepackage{graphicx}
\usepackage{cite}
\usepackage{color}
\usepackage{changes}
\usepackage{algorithm}
\usepackage{algorithmic}
\usepackage{acronym}
\usepackage{subcaption}
\usepackage{bm}
\usepackage{hyperref}
\hypersetup{
    colorlinks=true,
    allcolors=blue,
    }

\DeclareMathAlphabet{\mathbit}{OML}{cmr}{bx}{it}

\newacro{TDD}{time-division duplexing}
\newacro{CSI}{channel state information}
\newacro{DL}{downlink}
\newacro{UL}{uplink}
\newacro{BS}{base station}
\newacro{MS}{mobile station}
\acrodefplural{MS}{mobile stations}
\newacro{MSE}{mean square error}
\newacro{MMSE}{minimum mean square error}
\newacro{SVD}{singular value decomposition}
\newacro{AM}{alternating minimization}
\newacro{OFDM}{orthogonal frequency-division multiplexing}
\newacro{mmWave}{millimeter wave}
\newacro{OMP}{orthogonal matching pursuit}
\newacro{MIMO}{multiple-input multiple-output}
\newacro{RF}{radio frequency}
\newacro{LS}{least squares}
\newacro{MRC}{maximum ratio combiner}
\newacro{ZF}{zero-forcing}
\newacro{CS}{compressive sensing}
\newacro{ULA}{uniform linear array}
\newacro{ADC}{analog-to-digital converter}
\newacro{AoA}{angles-of-arrival}
\newacro{AoD}{angles-of-departure}
\newacro{CRLB}{Cram\'{e}r-Rao lower bound} 
\newacro{NMSE}{normalized mean squared error}
\newacro{CB}{coordinated beamforming}
\newacro{NMSE}{normalized mean-squared error}
\newacro{SINR}{signal-to-interference-plus-noise ratio}
\newacro{LLF}{log-likelihood function}
\newacro{SW-OMP}{simultaneous weighted - orthogonal matching pursuit}
\newacro{FDD}{frequncy-division duplexing}
\newacro{SS-SW-OMP+Th}{subcarrier-selection simultaneous weighted - orthogonal matching pursuit + thresholding}

\newcommand{\B}[1]{\mathbit{#1}}

\DeclareMathOperator{\Transpose}{T}
\DeclareMathOperator{\Hermitian}{H}
\newcommand{\Tr}{{\Transpose}}
\newcommand{\He}{{\Hermitian}}
\DeclareMathOperator{\Exp}{\mathbb{E}}
\DeclareMathOperator{\blockdiag}{blockdiag}

\DeclareMathOperator{\trace}{tr}
\DeclareMathOperator{\rank}{rank}
\DeclareMathOperator{\vect}{vec}

\newcommand{\Npu}{{N_{\text{p},u}}}

\newcommand{\Gr}{{G_\text{MS}}}
\newcommand{\Gt}{{G_\text{BS}}}
\newcommand{\Nr}{{N_\text{MS}}}
\newcommand{\NQ}{{N_\text{Q}}}
\newcommand{\Nt}{{N_\text{BS}}}
\newcommand{\Nru}{{N_{\text{MS},u}}}
\newcommand{\LBS}{{L_\text{BS}}}
\newcommand{\Lru}{{L_{\text{MS},u}}}
\newcommand{\Ns}{{N_{\text{s}}}}
\newcommand{\Nsu}{{N_{\text{s},u}}}
\newcommand{\MSE}{{\text{MSE}}}

\newcommand{\MMSE}{{\text{MMSE}}}
\newcommand{\DL}{{\text{DL}}}
\newcommand{\UL}{{\text{UL}}}
\newcommand{\RF}{{\text{RF}}}
\newcommand{\BB}{{\text{BB}}}

\newcommand{\eul}{{\text{e}}}
\newcommand{\imj}{{\text{j}}}
\newcommand{\PRF}{{\B{P}_\RF}}

\newcommand{\FRF}{{\B{F}_\RF}}
\newcommand{\FRFH}{{\B{F}_\RF^\He}}

\newcommand{\PBBu}{{\B{P}_\text{BB}^u}}
\newcommand{\PBBi}{{\B{P}_\text{BB}^i}}

\newcommand{\WRFu}{\B{W}_\RF^u}
\newcommand{\WRFuH}{\B{W}_\RF^{u,\He}}
\newcommand{\WBBu}{\B{W}_\text{BB}^u}
\newcommand{\WBBuH}{\B{W}_\text{BB}^{u,\He}}
\newcommand{\FBBu}{\B{F}_\text{BB}^u}
\newcommand{\FBBuH}{\B{F}_\text{BB}^{u,\He}}
\newcommand{\TBBu}{\B{T}_\text{BB}^u}
\newcommand{\TRFu}{\B{T}_\text{RF}^u}

\newcommand{\real}{\mathfrak{Re}}

\DeclareMathOperator{\diag}{diag}

\DeclareMathOperator{\blkdiag}{blkdiag}

\DeclareMathOperator{\Frob}{F}

\usepackage{tikz}
\usetikzlibrary{calc}

\newcommand{\positiontextbox}[4][]{%
  \begin{tikzpicture}[remember picture,overlay]
    \node[inner sep=3pt, fill=yellow,align=left,draw,line width=1pt,#1] at ($(current page.north west) + (#2,-#3)$) {\parbox{.95\paperwidth}{#4}};
  \end{tikzpicture}%
}
\begin{document}

%DISCLAIMER
%%%%%%%%%%%%%%%%%%%%%%%%%%%%%%%%%%%%%%%%%%%%%%%%%%%%%%%%
\onecolumn
\begingroup

\setlength\parindent{0pt}
\fontsize{14}{14}\selectfont

\vspace{1cm} 
\textbf{This is an ACCEPTED VERSION of the following published document:}

\vspace{1cm} 
J. P. González-Coma, J. Rodríguez-Fernández, N. González-Prelcic, L. Castedo, y R. W. Heath,``Channel Estimation and Hybrid Precoding for Frequency Selective Multiuser mmWave MIMO Systems'', \textit{IEEE J. Sel. Top. Signal Process.}, vol. 12, no. 2, pp. 353-367, May 2018, doi: 10.1109/JSTSP.2018.2819130

\vspace{1cm} 
Link to published version: \url{https://doi.org/10.1109/JSTSP.2018.2819130}

\vspace{3cm} 

\textbf{General rights:}

\vspace{1cm} 
\textcopyright 2018 IEEE. This version of the article has been accepted for publication, after peer review. \href{https://creativecommons.org/licenses/by-nc-nd/4.0/}{Personal use of this material is permitted. Permission from IEEE must be obtained for all other uses, in any current or future media, including reprinting/republishing this material for advertising or promotional purposes, creating new collective works, for resale or redistribution to servers or lists, or reuse of any copyrighted component of this work in other works.}
\twocolumn
\endgroup
\clearpage
%%%%%%%%%%%%%%%%%%%%%%%%%%%%%%%%%%%%%%%%%%%%%%%%%%%%%%%%%%%%
%%%%%%%%%%%%%%%%%%%%%%%%%%%%%%%%%%%%%%%%%%%%%%%%%%%%%%%%%%%%

\title{Channel estimation and hybrid precoding for frequency selective multiuser mmWave MIMO systems}
\author{\IEEEauthorblockN{Jos\'e P. Gonz\'alez-Coma\IEEEauthorrefmark{1},~Javier Rodr\'iguez-Fern\'andez\IEEEauthorrefmark{2},~Nuria Gonz\'alez-Prelcic\IEEEauthorrefmark{2},\\~Luis Castedo\IEEEauthorrefmark{1}~and~Robert W. Heath Jr.\IEEEauthorrefmark{3}}\\
	\IEEEauthorblockA{%
		\IEEEauthorrefmark{1}%
		Universidade da Coru\~na,~CITIC,
		Spain~
		\IEEEauthorrefmark{2}%
		Universidade de Vigo, Spain~\\
		\IEEEauthorrefmark{3}%
		The University of Texas at Austin, USA \\
		Email:\{\texttt{jose.gcoma,luis}\}\texttt{@udc.es},~\\\{\texttt{jrodriguez,nuria}\}\texttt{@gts.uvigo.es},~\texttt{rheath@utexas.edu}}  
\thanks{
	This work has been partially funded by Xunta de Galicia (ED431C 2016-045, ED341D R2016/012, ED431G/01, ED431G/04), AEI of Spain (TEC2015-69648-REDC, TEC2016-75067-C4-1-R, TEC2016-75103-C2-2-R),  ERDF funds (AEI/FEDER, EU) and the National Science Foundation under Grant No. 1702800.}
}

\maketitle

%%%%%%%%%%%%%%%%%%%%%%%%%%INCLUDING DISCLAIMER AT THE BOTTOM
\positiontextbox{10.75cm}{27cm}{\footnotesize \textcopyright 2018 IEEE. This version of the article has been accepted for publication, after peer review. Personal use of this material is permitted. Permission from IEEE must be obtained for all other uses, in any current or future media, including reprinting/republishing this material for advertising or promotional purposes, creating new collective works, for resale or redistribution to servers or lists, or reuse of any copyrighted component of this work in other works. Published version:
\url{https://doi.org/10.1109/JSTSP.2018.2819130}}
%%%%%%%%%%%%%%%%%%%%%%%%%%%%%%%%%%%%%%%%%%%%%%%%%%%%%%%%%%%%
%%%%%%%%%%%%%%%%%%%%%%%%%%%%%%%%%%%%%%%%%%%%%%%%%%%%%%%%%%%%

\begin{abstract}
Configuring the hybrid precoders and combiners in a \ac{mmWave} multiuser (MU) \ac{MIMO} system is challenging in frequency selective channels. In this paper, we develop a system that uses compressive estimation on the uplink to configure precoders and combiners for the \ac{DL}. In the first step, the \ac{BS} simultaneously estimates the channels from all the \acp{MS} on each subcarrier. To reduce the number of measurements required, compressed sensing techniques are developed that exploit common support on the different subcarriers. In the second step, exploiting reciprocity and the channel estimates, the base station designs hybrid precoders and combiners. Two algorithms are developed for this purpose, with different performance and complexity tradeoffs: 1) a factorization of the purely digital solution, and 2) an iterative hybrid design. Extensive numerical experiments evaluate the proposed solutions comparing to state-of-the-art strategies, and illustrating design tradeoffs in overhead, complexity, and performance.
 \end{abstract}

\section{Introduction}
Multiuser \ac{MIMO} communication is challenging in frequency selective millimeter wave communication systems. The main difficulties are a byproduct of the hybrid analog-digital array architecture, which splits \ac{MIMO} beamforming algorithms between analog and digital domains, to reduce overall power consumption \cite{AlMoGoHe14}. In such a system, the analog combining network prevents direct access to each antenna output, the SNR is low due to the large bandwidth, the channel dimensionality is high, and the analog combining is nominally frequency flat while the channel is frequency selective. This tends to require a lengthy training phase to estimate the channel \cite{AlkhaAyachLeusHeath}, a process which is aggravated in the multiuser setting, where channels are required for each active user.  Precoding and combining is also non-trivial because the analog portion is frequency flat with constrained values, and the number of \ac{RF} chains is limited. This requires designing multiple precoders and combiners subject to non-convex constraints, which makes it difficult to design optimal algorithms with low complexity. In this paper, we develop strategies for frequency selective channel estimation and precoder/combiner design in MU-\ac{MIMO} communication systems.

\subsection{Prior Work}

In this section, we review prior work on channel estimation and  precoder design in millimeter wave \ac{MIMO} systems, emphasizing  contributions that assume a frequency selective channel model.

% Review on channel estimation
Most of prior work on channel estimation at \ac{mmWave} focuses on narrowband channels \cite{AlkhaAyachLeusHeath},\cite{LeeGilLee},\cite{MendezRusuGonzalezAlkhaHeath}, while the \ac{mmWave} channel is indeed frequency selective. The limited work assuming a  \ac{mmWave} frequency-selective channel \cite{FreqChanEst17},\cite{ChanEst17Hybrid}, only considers the single user (SU) scenario. In multiuser settings, recent work focuses  on narrowband channel estimation \cite{ChEstmmWaveMUMIMOPARAFAC},\cite{MUPrecChEstHybridmmWave},\cite{CompressiveChEstmmWaveMUMIMOPilotReuse}. To the best of our knowledge, only \cite{GaommWaveMassiveMIMONearLOS} and \cite{GlobecomChesthybridpreccomb} address the problem of frequency-selective multiuser channel estimation at \ac{mmWave}. In \cite{GaommWaveMassiveMIMONearLOS}, the channels for the different users are estimated on the uplink, but the channels have a large Ricean factor thus only the strongest path for each user is estimated. In  \cite{GlobecomChesthybridpreccomb}, the channels are estimated by the MSs on the downlink and then the resulting \ac{CSI} is fed back to the \ac{BS}. The main limitations of this algorithm are that it does not exploit reciprocity and further it supposes that the channel directions lie on a specific grid. 

% Review on hybrid precoding; SU and MU, narrowband case

The idea of a hybrid analog-digital solution for the precoders and combiners in a \ac{MIMO} system was first proposed in \cite{ZhMoKu05}, and developed in \cite{AyRaAbPiHe14} for sparse \ac{mmWave} channels. A common approach to design the hybrid precoders/combiners is to factorize an all-digital solution into the analog and baseband domains. Several methods have been proposed to perform this factorization for the SU and narrowband case, most of them depending on the relationship between the number of streams and the number of \ac{RF} chains. Perfect factorization of the precoders in the downlink is obtained in \cite{ZhMoKu05}, but implies a large number of \ac{RF} chains. A sparse reconstruction problem with hardware constraints is formulated in \cite{AyRaAbPiHe14} to find the precoders, but results in high complexity and assumes perfect \ac{CSI} at the receiver side. Good approximations of the all-digital solution can be obtained with the greedy algorithm in \cite{MeRuGoHe15}, and fewer \ac{RF} chains. Other work such as \cite{RuMeGoHe16} proposes low complexity approaches to compute the precoders, but are also limited to  the SU and narrowband case. For the MU large antenna array setup, asymptotically optimal hybrid precoders were analyzed in \cite{PaAlHe16}. A joint solution for precoders and combiners in a MU scenario based on imperfect  \ac{CSI} was proposed in \cite{AlLeHe15} for frequency flat channels. Therein, the \ac{RF} precoders and combiners are found using beam training algorithms, while inter-user interference is canceled by baseband precoders. A similar approach was presented in \cite{DaCl17}. The \ac{RF} precoder is chosen from a codebook whereas the baseband counterpart is designed employing second-order channel statistics.  

% Review on hybrid precoding, wideband case, SU
The narrowband solutions for the design of the hybrid precoders and combiners cannot be, in general, extended to the wideband scenario. The main reason behind is that the \ac{RF} precoders and combiners are frequency flat, and have to be jointly designed for all the subcarriers, while the digital precoders/combiners are frequency selective.
While a few solutions have been proposed for the wideband scenario in the SU case \cite{PaAlHe16,YuShZhLe16}, all of them focus only on precoding without estimation. Considering imperfect  \ac{CSI} and a single user, a precoder design was proposed in  \cite{AlHe16}. Different search methods to find the \ac{RF} precoder from a codebook were proposed, while the baseband counterpart maximizes the mutual information.  The precoder design in \cite{AlHe16} though is only useful for the single user setting. 

Different precoding algorithms were proposed in \cite{YuShZhLe16} for the frequency selective single user case, where a search on the Riemannian manifold was performed to find the \ac{RF} precoders. The complexity of the  alternating minimization (AM) algorithm proposed there is prohibitive in our scenario, since the space search scales with the number of streams and subcarriers, which is large in the MU frequency selective scenario. In addition, authors in \cite{YuShZhLe16} present a low complexity alternative by restricting the columns of the baseband precoders to be mutually orthogonal. The motivation behind this constraint does not apply for some metrics as, e.g. \ac{MSE} as considered in our paper. 

%Paragraph about  MU case, wideband
Two solutions have been proposed for the frequency selective MU case  \cite{BoLeHaVa16,ChWaLi17}.
The first work assumes perfect \ac{CSI}, only addresses the problem of precoder design,  and evaluates the solution only for the high SNR regime. The algorithm for the MU frequency selective scenario developed in  \cite{ChWaLi17} exhibits a good tradeoff performance-complexity, but it only works with one \ac{RF} chain at both the transmitter and the receiver.

To the best of our knowledge, a solution for the the joint precoder and combiner design for the \ac{mmWave} frequency selective MU scenario under imperfect \ac{CSI} is missing except for \cite{GoGoCaHe16}. In that work, the autocorrelation of the received signal and the cross-correlation between the transmitted and received signal are estimated to compute precoders and combiners based on \ac{MSE} metric. The iterative design in \cite{GoGoCaHe16} employs several transmissions in the downlink and the uplink to set precoders and combiners. The digital solutions are next factorized via \cite[HD-LSR Algorithm]{RuMeGoHe16} at each step. This results in accumulated performance losses. Furthermore, the large training overhead might be unrealistic when the time coherence of the channel is short, for example with high mobility \ac{MS}s.

\subsection{Contributions}
In this paper, we propose  solutions to the problems of channel estimation and hybrid design of the precoders and combiners for a multiuser multistream \ac{MIMO} system operating in a frequency selective \ac{mmWave} channel.   We  assume  \ac{OFDM} modulation, which is commonly employed to simplify equalization in frequency selective channels. Next, we summarize the specific contributions of our work:

\begin{itemize}
	\item We propose an approach for joint estimation of the MU frequency selective channels, leveraging the jointly-sparse structure of the channels for different \ac{MS}s and exploiting reciprocity.  Compared to \ac{MIMO} \ac{DL} channel estimation, the consequences of this choice are twofold: i) it unnecessarily increases feedback overhead for the \ac{BS} to design hybrid precoders and combiners, and ii) since the number of antennas at a \ac{MS} is expected to be significantly smaller than at the \ac{BS}, a more lengthy training stage is required to obtain channel estimates. Our design overcomes these limitations. In addition, previous work on MU \ac{mmWave} frequency selective channel estimation  only considers on-grid angular parameters \cite{GlobecomChesthybridpreccomb}.
	\item We show theoretically and numerically that with a hybrid architecture at the receiver and transmitter sides, MU \ac{UL} channel estimation at the \ac{BS}, exploiting reciprocity of \ac{TDD} systems,  outperforms \ac{DL} channel estimation at the \ac{MS}s for a fixed overhead, even neglecting feedback. This is well known in massive \ac{MIMO} with fully digital and hybrid architectures \cite{FlRuTuLaEd17}, but it is not obvious with compressive estimation. Further, we show that when the AoAs and AoDs fall within a quantized spatial grid, the estimation error lies very close to the \ac{CRLB}.
	\item We design \ac{UL} precoders at each \ac{MS} that maximize their individual mutual information. The design we propose for the \ac{UL} combiner at the \ac{BS} is based on several criteria, exploiting knowledge on the \ac{UL} precoder used by each \ac{MS}, such that inter-user interference is mitigated. This scheme only depends on the \ac{CSI} of each \ac{MS}, and presents advantages to perform hybrid factorizations.
	\item We propose two approaches to compute baseband and \ac{RF} filters: 1) factorization of a digital design into baseband and \ac{RF} counterparts, as done in \cite{ZhMoKu05,AyRaAbPiHe14,YuShZhLe16,BoLeHaVa16,AhKhSh17,PaGhDi16}; 2) similar to \cite{StRa15,AlLeHe15}, we propose an iterative solution based on AM to directly design the hybrid precoders and combiners.
	\item We numerically show the connections between the performance of the designed precoders and combiners and the quality of the channel estimates obtained by the proposed algorithm. Extensive experiments show the good performance of our designs, compared to state of the art methods.
\end{itemize}

The paper is organized as follows: Section~\ref{sec:model} introduces the system model. Section~\ref{sec:problForm} formulates the problem for the design of precoders and combiners. In section~\ref{sec:hybridDesign}, different approaches to design the hybrid precoders and combiners are introduced. Section~\ref{sec:estimation} describes an algorithm for the estimation of multi-user frequency-selective channels. Section~\ref{sec:simulations} is devoted to the numerical experiments. Finally, Section~\ref{sec:conclusions} collects the main conclusions derived from this work.

%\textbf{Notation}:
%%%%%%%%%%%%%%%%%%%%%%%%%%%%%%%%%%%%%%%%%%%%%%%%%%%%%%%%%%%%%%%%%%%%%%%%%%%%%%%%%%%%%
\section{System Model}\label{sec:model}
%%%%%%%%%%%%%%%%%%%%%%%%%%%%%%%%%%%%%%%%%%%%%%%%%%%%%%%%%%%%%%%%%%%%%%%%%%%%
In this section, we introduce the models and assumptions for the different blocks of the communication system considered in this paper.

\paragraph{System block diagram}

We consider first the \ac{DL} of a \ac{MIMO} communication system operating at \ac{mmWave}, where a \ac{BS} equipped with $\Nt$ antennas transmits information to $U$ \ac{MS}s, each of them using $\Nru$ antennas. We assume that the number of \ac{RF} chains at the \ac{BS} and the \ac{MS}s are $\LBS<\Nt$ and $\Lru<\Nru$, which is realistic for practical scenarios. We consider a network of phase shifters connecting each antenna with all the \ac{RF} chains at both the transmitter an the receiver. Moreover, \ac{BS} allocates several independent data streams $\Nsu\leq \Lru$ to each \ac{MS}. The total number of data streams is assumed to be smaller than the number of \ac{RF} chains at the \ac{BS}, i.e., $\Ns=\sum_{u=1}^{U}\Nsu\leq \LBS$.

To deal with the frequency selective channel, \ac{OFDM} modulation is employed with a large enough cyclic prefix to avoid intercarrier interference. We denote by $\B{s}_u^\DL[k]$ the Gaussian signal data transmitted to \ac{MS} $u$ and subcarrier $k$ in the \ac{DL}. We assume the following statistical properties:  $\Exp[\B{s}^\DL_u[k]]=\mathbf{0}$, $\Exp[\B{s}_u^\DL[k]\B{s}_u^{\DL,\He}[k]]=\mathbf{I}_{\Nsu}$ and $\Exp[\B{s}_u^\DL[k]\B{s}_j^{\DL,\He}[k]]=\mathbf{0}$ for $j\neq u$. The symbol vectors are linearly processed using the hybrid precoder $\B{P}_u[k]=\PRF\PBBu[k]\in\mathbb{C}^{\Nt\times \Nsu}$ resulting from the product of the baseband and the \ac{RF} precoders, $\PBBu[k]\in\mathbb{C}^{\LBS\times \Nsu}$ and $\PRF\in\mathbb{C}^{\Nt\times \LBS}$. At the \ac{MS} end, the received signal is linearly filtered with the hybrid combiner $\B{W}_u[k]=\WRFu\WBBu[k]$, with $\WRFu\in\mathbb{C}^{\Nru\times \Lru}$ and $\WBBu[k]\in\mathbb{C}^{\Lru\times \Nsu}$. Since the \ac{RF} precoders and combiners are implemented using analog phase shifters, their entries are restricted to have constant modulus $|[\PRF]_{i,j}|^2=1$, $|[\WRFu]_{m,n}|^2=1$. 

Channel estimation is performed at the \ac{BS} during a training phase prior to data transmission. In particular, the \ac{BS} simultaneously estimates the channels for all the \ac{MS}. The assumption of channel reciprocity between the \ac{DL} and the \ac{UL} of \ac{TDD} systems is not impractical when hardware calibration is considered (see strategies proposed in \cite{BjLaMa16,HaCa17} and references therein). We assume the \ac{BS} can feedforward its channel estimate to the $i$-th \ac{MS}. Accordingly, the \ac{CSI} estimate at each \ac{MS} matches that in the \ac{BS}. No further information is shared to design precoders and combiners. 

\paragraph{Channel model}
Let us introduce the $D$-delay channel model for the link between the \ac{BS} and the \ac{MS} $u$.  The $d$-th delay tap, $\B{H}_{u,d}\in\mathbb{C}^{\Nru\times \Nt}$,  follows the expression \cite{PaAlHe16}
\begin{equation}
\B{H}_{u,d}=\gamma\sum_{p=1}^{\Npu}\alpha_{u,p}p_{\text{rc},u}(dT_s-\tau_{u,p})\B{a}_{\text{MS},u}(\theta_{u,p})\B{a}_{\text{BS}}^\He(\phi_{u,p}),
\label{eq:channelModel}
\end{equation}	
where $\Npu$ is the number of channel paths,  $\gamma=\sqrt{\Nt\Nru/\Npu}$ with the total number of receive antennas $\Nr=\sum_{u=1}^U \Nru$,   $\B{a}_{\text{MS},u}(\theta_{u,p})\in\mathbb{C}^{\Nru}$ and $\B{a}_{\text{BS}}(\phi_{u,p})\in\mathbb{C}^\Nt$ are the array response vectors of the receiver and the transmitter, $p_{\text{rc},u}(t)$ denotes the combined effects of pulse shaping and analog filtering  and $T_s$, $\tau_{u,p}$ and $\alpha_{u,p}$  are the sampling interval, the delay, and the gain. In the frequency domain, the channel response is
\begin{equation}
\B{H}_u[k]=\sum_{d=1}^{D}\B{H}_{u,d}\eul^{-\imj\frac{2\pi kd}{K}}.
\label{eq:channelModelFreq}
\end{equation}
An equivalent representation of the uplink channel is given by
\begin{equation}
\B{H}_u^\He[k]=\B{A}_{\text{BS}_u}\B{\Delta}_{u}[k]\B{A}_{\text{MS}_u}^\He,
\label{eq:equivCh}
\end{equation}
with the matrices $\B{A}_{\text{BS}_u}\in\mathbb{C}^{\Nt\times \Npu}$ and $\B{A}_{\text{MS}_u}\in\mathbb{C}^{\Nru\times \Npu}$ containing the steering vectors for the \ac{BS} and the \ac{MS} evaluated at the \ac{AoA} and \ac{AoD}, and the diagonal matrix $\B{\Delta}_{u}[k]\in\mathbb{C}^{\Npu\times \Npu}$ containing the gains for each of the $\Npu$ paths. 

Using the extended virtual channel model \cite{HeGoRaRoSa16}, the channel matrix in \eqref{eq:equivCh} can be approximated as
\begin{equation}
\B{H}_u^\He[k] \approx \tilde{\B{A}}_\text{BS} \B{\Delta}_u^\text{v}[k] \tilde{\B{A}}_{\text{MS},u}^\He,
\label{eq:extended_channel}
\end{equation}
where $\B{\Delta}_u^\text{v}[k] \in \mathbb{C}^{G_\text{BS} \times G_\text{MS}}$ is a sparse matrix containing the channel gains of the quantized spatial frequencies in its non-zero elements. The dictionary matrices $\tilde{\B{A}}_\text{BS} \in \mathbb{C}^{N_\text{BS} \times G_\text{MS}}$, $\tilde{\B{A}}_{\text{MS},u} \in \mathbb{C}^{N_{\text{MS},u} \times G_\text{MS}}$ contain the array response vectors for the \ac{BS} and $u$-th \ac{MS} evaluated on spatial grids of sizes $G_\text{BS}$ and $G_\text{MS}$.

\paragraph{Downlink received signal model}
According to our assumption of the cyclic prefix length, the channel in the frequency domain is decomposed into $K$ equivalent narrow-band channels. For subcarrier $k$ and \ac{MS} $u$, the received signal in the downlink is given by
\begin{align}
\B{x}_u[k]=\B{H}_u[k]\sum\nolimits_{i=1}^{U}\PRF\PBBi[k]\B{s}_i^\DL[k]+\B{n}_u[k].
\label{eq:downx}
\end{align}
where $\B{n}_u[k]\sim\mathcal{N}_\mathbb{C}(\mathbf{0},\sigma_n^2\mathbf{I}_{\Nru})$ is the additive Gaussian noise (AWGN). It is important to highlight that the \ac{RF} precoders are frequency flat, i.e., they have to be jointly optimized for all the subcarriers. Analogously, the \ac{RF} combiners $\WRFu$ are also frequency flat. The output signal after combining at the \ac{MS} $u$ and subcarrier $k$ is
\begin{align}
\B{y}^{\DL}_u[k]=\WBBuH[k]\WRFuH\B{x}_u[k].
\label{eq:downshat}
\end{align}

\paragraph{Uplink received signal model} $\B{H}_u^\He[k]$ is the channel response in the frequency domain for the link between the \ac{MS} $u$ and the \ac{BS}. The baseband and \ac{RF} precoders and combiners are $\TBBu[k]\in\mathbb{C}^{\Lru\times \Nsu}$, $\TRFu\in\mathbb{C}^{\Nru\times \Lru}$ and $\B{F}_\RF\in\mathbb{C}^{\Nt\times\LBS}$, $\FBBu[k]\in\mathbb{C}^{\LBS\times \Nsu}$, and the noise is $\B{n}\sim\mathcal{N}_\mathbb{C}(\mathbf{0},\sigma_n^2\mathbf{I}_N)$. Thus, the received signal in the uplink at subcarrier $k$ is given by
\begin{equation}
\B{x}[k]=\sum_{i=1}^{U}\B{H}_i^\He[k]\B{T}^i_\RF\B{T}^i_\text{BB}[k]\B{s}_i^\UL[k]+\B{n}[k].
\label{eq:ULrecSig}
\end{equation} 
The received signal after hybrid combining is $\B{y}_u^\UL[k]=\FBBuH[k]\B{F}_\RF^\He\B{x}[k]$.

Taking into account the signal model in \eqref{eq:ULrecSig}, the \ac{BS} simultaneously estimates the $U$ frequency-selective channels $\B{H}_i^\He[k]$, $i = 1,\ldots,U$, $k = 0,\ldots,K-1$ following the procedure described in section \ref{sec:estimation}, such that channel estimates $\hat{\B{H}}_i^\He[k]$ are computed.

\paragraph{Performance metrics}
Assuming Gaussian signaling and linear precoding/combining, we establish the achievable sum-rate as the system performance metric, yielding \cite{AyRaAbPiHe14,GoJaJiVi03}
\begin{align}
\sum_{u=1}^{U}R_u&=\frac{1}{K}\sum_{u=1}^{U}\sum_{k=1}^{K}\log_2\det\left(
\mathbf{I}_{\Nsu}+\B{X}_u^{-1}[k]\right.\nonumber\\
&\left.\times
\B{W}_u^\He[k]\B{H}_u[k]\B{P}_u[k]\B{P}_u^{\He}[k]\B{H}_u^{\He}[k]\B{W}_u[k]\right),
\label{eq:sumrate}
\end{align}
with $\B{X}_u[k]=\sum_{i\neq u}\B{W}_u^\He[k]\B{H}_u[k]\B{P}_i[k]\B{P}_i^{\He}[k]\B{H}_u^{\He}[k]\B{W}_u[k]+\sigma_n^2\B{W}_u^\He[k]\B{W}_u[k]$. The interference is treated as noise. 

To evaluate the performance of the proposed channel estimation strategy, we use the normalized mean-squared error (NMSE),  which is defined as
\begin{equation}
\text{NMSE} = \frac{\sum_{u=1}^{U}{\sum_{k=1}^{K}{||\hat{\bm H}_u^\He[k] - \bm H_u^\He[k]||_F^2}}}{\sum_{u=1}^{U}{\sum_{k=1}^{K}{||\bm H_u^\He[k]||_F^2}}}.
\label{eq:NMSE}
\end{equation}

%%%%%%%%%%%%%%%%%%%%%%%%%%%%%%%%%%%%%%%%%%%%%%%%%%%%%%%%%%%%%%%%%%%%%%%%%%%%%%%%%%%%%
\section{Hybrid Precoder/Combiner Design}
\label{sec:problForm}
%%%%%%%%%%%%%%%%%%%%%%%%%%%%%%%%%%%%%%%%%%%%%%%%%%%%%%%%%%%%%%%%%%%%%%%%%%%%%%%%%%%%%
To design the hybrid precoders and combiners for the \ac{DL}, the desired goal would be finding a solution that maximizes the achievable sum-rate in \eqref{eq:sumrate} employing the available \ac{CSI}, i.e.,
\begin{align}
	&\max_{\PRF,\PBBu[k],\WRFu,\WBBu[k]}\sum_{u=1}^{U}R_u\,\text{s.t.}\, \|\PRF\PBBu[k]\|_{\Frob}^2=\frac{P_{\text{tx}}}{U\Nsu },\forall u,k\nonumber\\
	&\quad\text{and}\;|[\PRF]_{i,j}|^2=1,\forall i,j\;\; |[\WRFu]_{m,n}|^2=1,\forall m,n,u,\label{eq:problForm}
\end{align}
with a power constraint for each \ac{MS} and subcarrier, such that the total transmit power is $P_\text{tx}$. We also define the signal-to-noise ratio as $\text{SNR}=P_\text{tx}/(U\sigma_n^2)$ for $\Exp[\alpha_{u,p}]=1$ in \eqref{eq:channelModel}.

The problem formulation in \eqref{eq:problForm} is however difficult to tackle, even neglecting the constraints imposed by the phase shifters, due to the coupling between the precoders. In the following section we propose alternative optimization problems and solve them for an all-digital filter as a previous step to obtain the hybrid solution.
 
\subsection{Digital Design}
\label{sec:digitalDesign}
A number of solutions have been proposed for the design of precoders and combiners for the MU \ac{DL}, with the goal of removing the inter-user interference, e.g. \cite{SpSwHa04,PaWoNg04,ChShAnHe08}. Another family of solutions is based on \ac{MSE} dualities between the  \ac{DL} and the \ac{UL} \cite{BoVa12,GoJoCaCaSP16,GoGrJoCaTSP}. That is, to calculate the combiners in the dual \ac{UL}, and finding scaling factors to obtain the \ac{DL} precoders. Leveraging the duality,  the \ac{UL} precoders and combiners can be designed in closed form to minimize the sum-\ac{MSE}. The sum-\ac{MSE} establishes a lower bound for the achievable sum-rate, and it gets tighter when \ac{MS}s have similar power constrains and average channel gains \cite{TaChSr11}. Hence, the \ac{MSE} criterion is a common choice when trying to maximize the achievable sum-rate \cite{ChAgCaCi08,ShScBo08TSP}.

We propose to design the \ac{UL} precoders in a selfish way using the available \ac{CSI} at the \ac{MS}, similar to coordinated transmit-receive processing in \cite{SpSwHa04}, i.e., 
\begin{equation}
	\B{T}_u[k]=[\B{V}_u[k]]_{1:\Nru,1:\Nsu}\B{\Lambda}_u[k],
	\label{eq:ULprecoder}
\end{equation}
where $\hat{\B{H}}^\He_u[k]=\B{U}_u[k]\B{\Sigma}_u[k]\B{V}^\He_u[k]$, and $\B{\Lambda}_u[k]$ is the usual water-filling solution \cite{Te99}.
 
Observe that we have neglected the inter-user interference in the design of the uplink precoders. This is motivated by the lack of knowledge of the channels of the rest of the \ac{MS}s and the \ac{UL} combiners employed at the \ac{BS}. On the contrary, the \ac{BS} has information of all \ac{MS}' channels and can unilaterally determine the precoders at the \ac{MS}s (cf. \eqref{eq:ULprecoder}). Thus, for the uplink combiners we focus on minimizing the \ac{MSE}. The \ac{MSE} for \ac{MS} $u$ and subcarrier $k$ in the uplink is given by 
$\MSE_u^\UL[k]=\Exp[\|\B{s}_u^\UL[k]-\B{y}_u^\UL[k]\|^2_2]$. Expanding the expectation and simplifying
\begin{align}
&\MSE^\UL_u[k]=\Nsu-2\real\left\{\trace\left(\B{F}_u^\He
[k]\B{H}_u^\He[k]\B{T}_u[k]\right)\right\}\label{eq:MSEULCorr}\\
&+\trace\left(\B{F}_u^\He[k]\left(\sum_{i=1}^U\B{H}_i^\He[k]\B{T}_i[k]\B{T}_i^\He[k]\B{H}_i[k]+\sigma_n^2\mathbf{I}\right)\B{F}_u[k]\right).\nonumber
\end{align}
Note from \eqref{eq:MSEULCorr} that the  combiners are decoupled and can be independently designed for each \ac{MS} and subcarrier. The \ac{MMSE} combiner \cite{Wi64,VeVe10} is 
\begin{align}
\B{F}^u_\MMSE[k]=&\left(\sum_{i=1}^U\hat{\B{H}}_i^\He[k]\B{T}_i[k]\B{T}_i^\He[k]\hat{\B{H}}_i[k]+\sigma_n^2\mathbf{I}\right)^{-1}\nonumber\\
&\times\hat{\B{H}}_u^\He[k]\B{T}_u[k].\label{eq:DLcomb}
\end{align}
Provided that $\hat{\B{H}}_u[k]$ and consequently $\B{T}_u[k]$ are known at the \ac{BS}, we obtain \eqref{eq:DLcomb} for the \ac{UL} combiners.

Taking into account the characteristics of \ac{mmWave} channels, it is intriguing to evaluate the performance of other schemes for lower frequencies. In particular, \ac{MRC} strategy presents low complexity and good performance in the low SNR regime; \ac{CB} removes the inter-user interference, and it is asymptotically optimal for the achievable sum rate \cite{SpSwHa04,PaGhDi16}. \ac{MRC}  maximizing the SNR of the received signal is then \cite{BjBeOt14} 
\begin{equation}
\B{F}_\text{MRC}^{u}[k]=\hat{\B{H}}_u^\He[k]\B{T}_u[k].
\label{eq:FMRC}
\end{equation}
   
According to the equivalent channel model \eqref{eq:equivCh}, the inter-user interference matrix is
\begin{align}
\B{R}_{\bar{u}}^\UL[k]&=\sum_{i\neq u}\B{A}_{\text{BS}_i}\B{\Delta}_{i}[k]\B{A}_{\text{MS}_i}^\He\B{T}_i[k]\B{T}_i^\He[k]\B{A}_{\text{MS}_i}\B{\Delta}_{i}^\He[k]\B{A}_{\text{BS}_i}^\He.\nonumber
\end{align}
Observe that the rank of this matrix is bounded, that is $\rank(\B{R}_{\bar{u}}^\UL[k])\leq\sum_{i\neq u}\min(N_{\text{p},i},N_{\text{s},i})\leq\LBS-\Nsu$, according to the assumptions in our setup. Similar to \cite{SpSwHa04}, we propose to point the combiners in the direction of the nullspace of $\hat{\B{R}}_{\bar{u}}^\UL[k]$. Therefore, we obtain the \ac{CB} combiner as 
\begin{equation}
\B{F}^{u}_{\text{CB}}[k]=\B{U}_u[k]\B{U}_u[k]^\He\hat{\B{H}}_u^\He[k]\B{T}_u[k].
\label{eq:FZF}
\end{equation}
with $v=\rank(\hat{\B{R}}_{\bar{u}}^\UL[k])$ and $\B{U}_u[k]\in\mathbb{C}^{\Nt\times \Nt-v}$ being the basis of $\hat{\B{R}}_{\bar{u}}^\UL[k]$ nullspace.

%%%%%%%%%%%%%%%%%%%%%%%%%%%%%%%%%%%%%%%%%%%%%%%%%%%%%%%%%%%%%%%%%%%%%%%%%%%%%%%%%%%%%
\subsection{Hybrid Design}
%%%%%%%%%%%%%%%%%%%%%%%%%%%%%%%%%%%%%%%%%%%%%%%%%%%%%%%%%%%%%%%%%%%%%%%%%%%%%%%%%%%%%
\label{sec:hybridDesign}
In this section, we propose different approaches to design hybrid precoders and combiners. In Sec. \ref{sec:DDFac} we present an algorithm to solve the hybrid factorization of digital schemes. Moreover, in Sec. \ref{sec:AM} we directly compute the hybrid precoders and combiners by means of an \ac{AM}.

%%%%%%%%%%%%%%%%%%%%%%%%%%%%%%%%%%%%%%%%%%%%%%%%%%%%%%%%%%%%%%%%%%%%%%%%%%%%%%%%%%%%%
\subsubsection{Factorization of Digital Designs}
\label{sec:DDFac}
In this section, we consider the factorization of the digital solution obtained in Sec. \ref{sec:digitalDesign}. As in prior work, we use the Frobenius norm as the means of computing the distance between the unconstrained and the hybrid solution. This was shown in \cite{AyRaAbPiHe14} to be meaningful for precoding with a unitary constraint; low complexity approaches to the computation were devised in \cite{RuMeGoHe16}.
Minimizing the Frobenius norm is also meaningful for the \ac{MSE}, as shown in Appendix \ref{ap:Mseproof}. Unlike \cite{LeLe14,YuShZhLe16} where a single user is considered, the \ac{MS}s and the subcarriers are coupled in $\FRF$. By introducing the matrices $\B{F}_\MMSE=[\B{F}^1_{\MMSE}[1],\ldots,\B{F}^1_{\MMSE}[K],\ldots,\B{F}^U_{\MMSE}[K]]\in\mathbb{C}^{\Nt\times K\Ns}$ and $\B{F}_\BB=[\B{F}^1_{\text{BB}}[1],\ldots,\B{F}^1_{\text{BB}}[K],\ldots,\B{F}^U_{\text{BB}}[K]]\in\mathbb{C}^{\LBS\times K\Ns}$, the factorization problem becomes
\begin{align}
\min_{\FRF,\B{F}_\BB}&\|\B{F}_\MMSE-\FRF\B{F}_\BB\|_{\Frob}^2\;\text{s.t.}\;|[\FRF]_{i,j}|^2=1,\,\forall i,j\nonumber\\ &\text{and}\;\|\FRF\FBBu[k]\|_{\Frob}^2=\|\B{F}^u_{\MMSE}[k]\|_{\Frob}^2,\,\forall u,k. \label{eq:subpr2Ref}
\end{align}
An analogous formulation is obtained for the \ac{UL} precoders, but the factorization is individually performed for each \ac{MS}. Since the ideas for the \ac{UL} combiners apply for the \ac{UL} precoders, we focus on the hybrid design of the \ac{UL} combiners.

From the proposed methods in the literature, only those allowing $\LBS\ll K\Ns$ apply for the problem formulation in \eqref{eq:subpr2Ref}. The convex relaxation over the set of complex matrices with unitary entries presented in \cite[HD-LSR Alg.]{RuMeGoHe16} can be extended to the frequency selective scenario \cite{GoGoCaHe16}, and its performance will be evaluated in our experiments. It consist of finding the \ac{RF} precoders performing a search over a ball of radius $1+\beta$, where $\beta$ establishes a compromise between the search space size and the relaxation precision. Then, the baseband precoders for such \ac{RF} update are calculated using \ac{LS}. After the \ac{AM} procedure, a final power normalization is performed to fulfill the power constraints. 

\paragraph{Projected Gradient Factorization}
\label{sec:PG}

To design the \ac{UL} baseband combiners for given \ac{UL} \ac{RF} combiners, the main approach in the literature is to use the \ac{LS} solution.  That is, the \ac{UL} baseband combiner $\B{F}_\BB$ is the solution of the minimization problem
\begin{equation}
\min_{\B{F}_\BB}\|\B{F}_\MMSE-\B{F}_\RF\B{F}_\BB\|_{\Frob}^2,
\label{eq:LSproblem}
\end{equation}
which is given by the closed-form expression $\B{F}_\BB=\B{F}_\RF^\dagger\B{F}_\MMSE$. This solution, however, has to be properly scaled to ensure that the power constraints are satisfied. Therefore, we compute the \ac{UL} baseband combiner as 
\begin{equation}
	\B{F}_\BB=\B{F}_\RF^\dagger\B{F}_\MMSE\tilde{\B{\Theta}},
	\label{eq:bbcombiner}
\end{equation}
 $\tilde{\B{\Theta}}=\blockdiag(\theta_{1}[k]\mathbf{I}_{N_{\text{s},1}},\ldots,\theta_{1}[K]\mathbf{I}_{N_{\text{s},1}},\ldots,\theta_{U}[K]\mathbf{I}_{N_{\text{s},U}})$ contains the factors resulting from the power normalization $\theta_u[k]=\sqrt{P_\text{tx}/(U\Nsu)}/\|\FRF\FBBu[k]\|_{\Frob}$. Substituting \eqref{eq:bbcombiner} in \eqref{eq:LSproblem} we obtain the distortion metric $d$
\begin{equation}
d=\|\B{F}_\MMSE\|_{\Frob}^2-\trace\big(\B{\Theta}\B{F}_\MMSE^\He\FRF(\FRFH\FRF)^{-1}\FRFH\B{F}_\MMSE\big),
\label{eq:distor}
\end{equation}
with $\B{\Theta}=2\tilde{\B{\Theta}}-\tilde{\B{\Theta}}^2$.  Note that
$\FRF$ is a tall matrix, and  $\B{\Pi}=\FRF(\FRFH\FRF)^{-1}\FRFH$ is an orthogonal projector. It is apparent that the accuracy of the \ac{LS} solution depends on the relationship between the rank of the digital solution and the  number of \ac{RF} chains $\LBS$. For example, in low-frequency architectures deploying an \ac{RF} chain per antenna we have $\rank(\B{F}_\MMSE)\leq\LBS=\Nt$. Hence, for any full rank $\FRF$ the projector becomes $\B{\Pi}=\mathbf{I}_\Nt$ and perfect factorization is ensured. 

Taking into account the distortion $d$ introduced by the \ac{LS} baseband filters and the power normalization \eqref{eq:distor}, we propose an alternative to the methods studied in \cite{RuMeGoHe16} which does not rely on \ac{AM} to design the \ac{RF} and baseband combiners. Indeed, the baseband combiners are implicitly updated in \eqref{eq:distor}, and the expression depends only on the \ac{RF} combiners. 

Now, we propose to design the \ac{RF} combiners by means of a projected gradient method. The objective function is $d$ in \eqref{eq:distor}, and $|[\FRF]_{i,j}|^2=1$ are the constraints defining the feasible set of solutions. First, let us split the distortion up for the \ac{MS}s and subcarriers as $d=\sum_{u=1}^{U}\sum_{k=1}^{K}d_{u,k}$,
\begin{equation}
d_{u,k}=\|\B{F}^{u}_\MMSE[k]\|_{\Frob}^2-2\sqrt{\frac{P_\text{tx}}{U\Nsu}}\|\B{A}_u[k]\|_{\Frob}+\frac{P_\text{tx}}{U\Nsu}
\label{eq:distDecomp}
\end{equation}
where $\B{A}_u[k]=(\FRFH\FRF)^{-1/2}\FRFH \B{F}^{u}_\MMSE[k]$. Thus, we pose the optimization problem as follows
\begin{equation}
	\min_{\FRF}\sum_{u=1}^{U}\sum_{k=1}^{K}d_{u,k}\quad\text{s.t.}\quad |[\FRF]_{i,j}|^2=1,\forall i,j.
\end{equation}
This problem is non-convex because of the analog restrictions. However, it is possible to reach a local optimum following the direction of the gradient, and projecting the solution to the feasible set afterwards \cite{Be99}. The gradient step obeys
\begin{equation}
\bar{\B{F}}_\RF = \FRF+s\frac{\partial d}{\partial \B{F}_\RF^*},
\label{eq:gradStep}
\end{equation} 
with the diminishing step size $s$ \cite{Be99}. The gradient in \eqref{eq:gradStep} is computed in Appendix \ref{ap:gradient} as
\begin{align}
\frac{\partial d}{\partial \B{F}_\RF^*}&=\sum_{u=1}^{U}\sum_{k=1}^{K}\frac{\sqrt{P_\text{tx}/(U\Nsu)}}{\|\B{A}_u[k]\|_{\Frob}}\big(\FRF(\FRFH\FRF)^{-1}\FRFH-\mathbf{I}_{\Nt}\big)\nonumber\\&\times\B{F}^{u}_\MMSE[k]\B{F}^{u,\He}_\MMSE[k]\FRF(\FRFH\FRF)^{-1}.
\label{eq:gradient}
\end{align}
Note that for $\LBS=\Nt$ the gradient is zero for full rank $\FRF$. Alternatively, the necessary condition for optimality is reached if all $\B{F}^{u}_\MMSE[k]$ are contained in the span of the \ac{RF} combiner.

As aforementioned, the \ac{RF} precoders do not satisfy the unitary entry constraint after the update. Hence, at each iteration $\bar{\B{F}}_\RF$ is projected to the feasible set by simply taking the phases 
\begin{equation}
\FRF =\arg(\bar{\B{F}}_\RF).
\label{eq:projStep}
\end{equation} 

\begin{algorithm}[!t]
	\caption{Hybrid Design by Projected Gradient (HD-PG)}
		\label{alg:ProjGrad}
	\begin{algorithmic}[1]
		\STATE Initialize: $\ell \leftarrow 0$, random $\FRF(\ell)$
		\REPEAT 
		\STATE $\ell \leftarrow \ell+1$
		\STATE $s\leftarrow s_0$
		\REPEAT 
		\STATE $\bar{\B{F}}_\RF\leftarrow \FRF(\ell-1)+s\frac{\partial d}{\partial \FRF^*}$
		\STATE$\FRF(\ell)\leftarrow\arg(\bar{\B{F}}_\RF)$
		\STATE $s\leftarrow\frac{s}{2}$
		\UNTIL {$d(\FRF(\ell-1))\geq d(\FRF(\ell))$}
		\UNTIL{$d(\FRF(\ell-1))-d(\FRF(\ell)) < \delta \;\text{or}\;\ell\geq \varepsilon$}
		\STATE $\B{F}\leftarrow\B{F}^{\RF,\dagger}(\ell)\B{F}_\MMSE$
		\FORALL {$u,l$ }
		\STATE $\FBBu[k]\leftarrow\FBBu[k]\times \sqrt{P_{\text{tx}}/(U\Nsu)}/\|\FRF(\ell)\FBBu[k]\|_{\Frob}$
		\ENDFOR
	\end{algorithmic}
\end{algorithm}

Algorithm \ref{alg:ProjGrad} summarizes the proposed method. Since every step of the algorithm reduces the error metric, which is bounded below by zero, the decreasing step size guarantees convergence to a local optimum \cite[Prop. 2.2.1]{Be99}. It is remarkable that, contrary to methods previously proposed where the power is merely normalized after algorithm's convergence, the power constraint is considered to compute the gradient. In other words, while the power constraints are transparent in Algorithm \ref{alg:ProjGrad}, their performance impact has not been determined for previous approaches. 

Algorithm \ref{alg:ProjGrad} can also be employed to find the \ac{UL} \ac{RF} precoders. However, they are individually designed for each \ac{MS}. Although the number of columns of the digital \ac{UL} precoding matrix is smaller than that of the \ac{UL} combining matrix, the number of subcarriers times the number of streams is still large. This fact makes the strategies assuming $\Lru\geq K\Nsu$ hardly applicable in the frequency selective scenario.

Observe that the flexibility of the HD-PG method makes it possible its utilization to factorize the precoder and combiners for any number of \ac{MS}s, streams, or subcarriers.

%%%%%%%%%%%%%%%%%%%%%%%%%%%%%%%%%%%%%%%%%%%%%%%%%%%%%%%%%%%%%%%%%%%%%%%%%%%%%%%%%%%%%
\subsubsection{Hybrid Design by Alternating Minimization}
\label{sec:AM}
In this section we propose ad hoc solutions directly incorporating the hybrid constraint in the design, as in \cite{YuShZhLe16,RuMeGoHe16}. The main advantage is that the computationally complex factorization is
avoided. This family of solutions consider a fixed \ac{RF} precoder/combiner already designed  according to some criterion. Next, the baseband counterpart is computed to optimize the same or a different criterion taking into account the \ac{RF} part previously obtained. Now, since the \ac{RF} optimization usually depends on the baseband matrices, the \ac{RF} and baseband precoders/combiners are alternatively updated until convergence. 

We start designing the \ac{UL} hybrid precoders by maximizing the mutual information, as in the digital solution in Sec. \ref{sec:digitalDesign}. 
Since the \ac{RF} precoder is common to all subcarriers, it has to be jointly designed for all of them. Let us define $\B{H}_u=[\B{H}_u^*[1],\ldots,\B{H}_u^*[K]]^\Tr$ containing all the channels for all the subcarriers. According to recent work, the common support of the frequency selective \ac{mmWave} channel model in \eqref{eq:channelModel} can be exploited to design the precoders/combiners \cite{VeGoHe16}. Likewise, we propose to use the right singular vectors corresponding to the largest $\Lru$ singular values as columns of the \ac{UL} precoder, with the \ac{SVD}  $\B{H}_u=\B{U}_{\B{H}_u}\B{\Sigma}_{\B{H}_u}\B{V}_{\B{H}_u}^\He$. The motivation behind this choice is to simultaneously increase the SNR for all the subcarriers. These vectors, however, have to be entry-wise normalized to satisfy the analog restrictions,  
\begin{equation}
	\B{T}_\RF^u=\arg\big([\B{V}_{\B{H}_u}]_{1:\Nru,1:\Lru}\big)
	\label{eq:AMprecoders}.
\end{equation}

For the \ac{RF} precoders in \eqref{eq:AMprecoders}, we compute the equivalent channels $\tilde{\B{H}}_u[k]=\B{H}_u^\He[k]\B{T}_\RF^u$. Next, the baseband precoding matrices $\B{T}_\BB^u[k]$ are computed as the mutual information maximizers of a \ac{MIMO} point-to-point link with the equivalent channel $\tilde{\B{H}}_u[k]$ \cite{Te99} [cf. \eqref{eq:ULprecoder}]. Again, this \ac{UL} hybrid precoding matrix can be determined at the \ac{BS} using only the channel estimates, meaning that the \ac{UL} precoders can be determined at the \ac{BS} for computing the \ac{UL} combiners.

Let us define the interference-plus-noise matrix in the uplink, $\B{R}_\text{I}^\UL[k]$, and the desired signal matrix, $\B{R}_{\text{D},u}^\UL[k]$, as
\begin{align}
&\B{R}_\text{I}^\UL[k]=\sum_{i=1}^{U}\B{H}_{i}^\He[k]\B{T}_\RF^i\B{T}_\BB^i[k]\B{T}_\BB^{i,\He}[k]\B{T}_\RF^{i,\He}\B{H}_i[k]+\sigma_n^2\mathbf{I}_\Nt,\nonumber\\
&\B{R}_{\text{D},u}^\UL[k]=\B{H}_{u}^\He[k]\B{T}_\RF^u\B{T}_\BB^u[k].
\label{eq:correlations}
\end{align}
Using this compact notation, the \ac{MSE}  considering the \ac{RF} and the baseband combiners is
\begin{align}
\MSE_u^\UL[k]&=N_{\text{s},u}-2\real\left\{\trace(\FBBuH[k]\B{F}_\RF^\He\B{R}_{\text{D},u}^\UL[k])\right\}\nonumber\\
&+\trace\left(\FBBuH[k]\B{F}_\RF^\He\B{R}_\text{I}^\UL[k]\B{F}_\RF\FBBu[k]\right).\label{eq:MSEULhybr}
\end{align} 
For the combiner in the uplink we propose a different criterion design as that of the precoder. In particular, the \ac{RF} combiner aims at maximizing the second term of \eqref{eq:MSEULhybr} for all \ac{MS}s and subcarriers, i.e., maximizing the SNR of the received signal 
\begin{equation}
\B{F}_\RF=\arg\left(\sum_{k=1}^{K}\sum_{u=1}^{U}\hat{\B{R}}_{\text{D},u}^\UL[k]\FBBuH[k]\right),
\label{eq:RFupdateAO}
\end{equation}
where $\hat{\B{R}}_{\text{D},u}^\UL[k]$ is the desired signal matrix in \eqref{eq:correlations} with the channel estimate $\hat{\B{H}}_u^\He[k]$.
This way, we have a closed-form expression for the \ac{RF} combiner for which the baseband combiners are considered to be given. The baseband combiners $\FBBu[k]$ are independently designed for each \ac{MS} and subcarrier for given \ac{RF} combiners, according to the \ac{MSE} criterion
\begin{equation}
\B{F}_\BB^{u,\MMSE}[k]=\left(\B{F}_\RF^\He\hat{\B{R}}^\UL_\text{I}[k]\B{F}_\RF\right)^{-1}\B{F}_\RF^\He\hat{\B{R}}_{\text{D},u}^\UL[k],
\label{eq:FBBMMSEAO}
\end{equation}
with $\hat{\B{R}}_{\text{D},u}^\UL[k]$ being the interference-plus-noise matrix \eqref{eq:correlations} for the estimated channel $\hat{\B{H}}_u^\He[k]$. Equations \eqref{eq:FBBMMSEAO} and \eqref{eq:RFupdateAO} motivate the \ac{AM} process. Thereby, the baseband and the \ac{RF} combiners are alternatively updated until convergence. Unfortunately, convergence cannot be theoretically guaranteed. Since the updates of the \ac{RF} and baseband combiners obey different criterion, it is possible that the cost function does not reduce at every step. However, we observe good results from numerical experiments.  

In Section \ref{sec:digitalDesign} we posed alternatives to the digital \ac{MMSE} \ac{UL} combiner. Likewise, we propose different baseband combiner designs to include in the \ac{AM} procedure. That is, for the \ac{RF} combiner in \eqref{eq:RFupdateAO} we employ \ac{MRC}-like and \ac{CB}-like baseband combiners instead of \eqref{eq:FBBMMSEAO} 
\begin{align}
\B{F}_\BB^{u,\text{MRC}}[k]&=\left(\B{F}_\RF^\He\B{C}_{\B{n}}\B{F}_\RF\right)^\dagger\FRFH\hat{\B{R}}_{\text{D},u}^\UL[k],\;\text{and}\nonumber\\
\B{F}_\BB^{u,\text{CR}}[k]&=\B{U}_u[k]\B{U}^\He_u[k]\FRFH\hat{\B{R}}_{\text{D},u}^\UL[k].
\label{eq:FBBMRCZFAO}
\end{align}
where $\B{U}_u[k]$ is the basis of null space of $\FRFH\hat{\B{R}}_{\bar{u}}^\UL[k]\FRF$ (cf. \eqref{eq:FZF}). Note that the bound for the rank in Sec. \ref{sec:digitalDesign} holds also here $\rank(\FRFH\hat{\B{R}}_{\bar{u}}^\UL[k]\FRF)\leq \rank(\hat{\B{R}}_{\bar{u}}^\UL[k])$.

\section{Compressive Channel Estimation}
\label{sec:estimation}

This section is devoted to provide a solution for the problem of estimating the multi-user uplink \ac{mmWave} frequency-selective MIMO channel. In this work, unlike in \cite{GlobecomChesthybridpreccomb}, we propose an alternative to perform channel estimation at the BS in the uplink. We provide the mathematical model for simultaneous estimation of the $UK$ channel matrices $\B{H}_u^\He[k]$, $\forall u,k$, and use the \ac{SW-OMP} algorithm in \cite{FreqChanEst17} to perform this task. Furthermore, the \ac{CRLB} for the estimation of the multi-user channel is provided. Finally, we propose a reasonably fair criterion to compare uplink and downlink channel estimation. In the simulations section, we will show that the estimation performance in the \ac{BS} significantly outperforms downlink channel estimation, and will also compare the \ac{CRLB} for both scenarios.

\subsection{Uplink Compressive Channel Estimation}
We assume that the \ac{BS} and the $U$ \ac{MS}s undertake a training stage in which the different \ac{MS}s simultaneously send $M$ training frames to the \ac{BS} using OFDM signaling. For the transmission of the $m$-th training frame at subcarrier $k$, each \ac{MS} uses a precoder $\bm T_{\text{tr},u}^{(m)}[k] \in \mathbb{C}^{\Nru \times \Lru}$ to transmit a training symbol $\bm s_u^{(m)}[k]$, and the \ac{BS} process the received MIMO signal with a combiner $\bm F_\text{tr}^{(m)}[k] \in \mathbb{C}^{\Nt \times \LBS}$. Accordingly, the multi-user received signal at the \ac{BS} can be written as
\begin{equation}
\bm y^{(m)}[k] = \bm F_\text{tr}^{(m)\He}[k] \sum_{u=1}^{U}{\bm H_u^\He[k]\bm T_{\text{tr},u}^{(m)}[k]\bm s_u^{(m)}[k]} + \bm n^{(m)}[k],
\label{equation:rx_signal_frame}
\end{equation}
where $\bm y^{(m)}[k]$ is the $\LBS \times 1$ received signal during the $m$-th training step, and $\bm n^{(m)}[k] \in \mathbb{C}^{L_\text{BS} \times 1}$ is the received post-combining additive colored Gaussian noise vector with distribution $\bm n^{(m)}[k] \sim {\cal N}(\bm 0, \bm F_\text{tr}^{(m)\He}[k] \bm F_\text{tr}^{(m)}[k])$. The received signal in \eqref{equation:rx_signal_frame} can be further expressed as
\begin{align}
&\bm y^{(m)}[k] = \bm F_\text{tr}^{(m)\He}[k] \underbrace{\big[\begin{array}{ccc} \bm H_1^\He[k]  & \ldots & \bm H_U^\He[k] \\ \end{array}\big]}_{\B{H}^\He[k]} \times \\
&\times \underbrace{\big[\begin{array}{ccc} 
\bm s_1^{(m)\Tr}[k]\bm T_{\text{tr},1}^{(m)\Tr}[k] 
\ldots 
\bm s_U^{(m)\Tr}[k]\bm T_{\text{tr},U}^{(m)\Tr}[k] \\ \end{array}\big]^\Tr}_{\bm x[k]} + \bm n^{(m)}[k],\nonumber
\end{align}
which will be used as baseline to derive the relation between the received signal and the vectorized channel matrices for each \ac{MS}. If we apply the $\vect\{\cdot\}$ operator to the previous equation, it is possible to write
\begin{equation}
\bm y^{(m)}[k] = \left(\bm x^\Tr[k] \otimes \bm F_\text{tr}^{(m)\He}[k] \right) \vect\left\{\B{H}^\He[k]\right\} + \bm n^{(m)}[k].
\label{equation:vect_rx_frame}
\end{equation}

Now, using the extended virtual channel model in \eqref{eq:extended_channel}, the vectorized channel matrix $\bm H_u^\He[k]$ for the $u$-th \ac{MS} can be approximated as
\begin{equation}
\vect\{\bm H_u^\He[k]\} \approx \overbrace{\left(\tilde{\bm A}_{\text{MS},u}^* \otimes \tilde{\bm A}_\text{BS}\right)}^{\tilde{\bm \Psi}_u} \vect\{\tilde{\bm \Delta}_u[k]\},
\label{equation:approx_channel}
\end{equation}
with $\tilde{\bm \Psi}_u \in \mathbb{C}^{\Nt \Nru \times G_\text{BS} G_\text{MS}}$ the dictionary matrix for the $u$-th user. The expression in \eqref{equation:approx_channel} approximately holds if the grid sizes of the dictionary matrices are large enough \cite{HeGoRaRoSa16}. Typically, the choices $G_\text{BS} \geq 2\Nt$ and $G_\text{MS} \geq 2\Nru$ are considered to estimate the channel \cite{FreqChanEst17},\cite{ChanEst17Hybrid}.
Thereby, the approximation in \eqref{equation:approx_channel} can be plugged into  \eqref{equation:vect_rx_frame} to obtain
\begin{equation}
\bm y^{(m)}[k] \approx \overbrace{\left(\bm x^\Tr[k] \otimes \bm F_\text{tr}^{(m)\He}[k] \right)}^{\bm \Phi^{(m)}[k]} \left(\bigoplus_{u=1}^{U}{\tilde{\bm \Psi}_u}\right) 
{\bm h}_\text{MU}^\text{v}[k] + \bm n^{(m)}[k],
\label{equation:vect_rx_frame_channel}
\end{equation}
with the vector ${\bm h}_\text{MU}^\text{v}[k]=[\vect\{{\bm \Delta}_1^{\text{v}}[k]\}^\Tr\ldots \vect\{{\bm \Delta}_U^{\text{v}}[k]\}^\Tr]^\Tr$ containing the concatenation of the vectorized $U$ sparse matrices $\bm \Delta_u^\text{v}[k]$, and the matrix $\bm \Phi^{(m)}[k] \in \mathbb{C}^{\LBS \times \Nt\sum_{u=1}^{U}{\Nru}}$ is termed the measurement matrix for channel reconstruction. 

The signal model in \eqref{equation:vect_rx_frame_channel} considers the general use of frequency-selective hybrid precoders and combiners to estimate the channel. However, the task of compressive channel estimation has computational complexity growing with the number of subcarriers $K$ \cite{FreqChanEst17}.  
To reduce this complexity, we consider the use of frequency-flat training precoders and combiners. Hence, $\bm T_{\text{tr},u}^{(m)}[k] = \bm T_{\text{tr},u}^{(m)}$, and $\bm F_{\text{tr}}^{(m)}[k] = \bm F_{\text{tr}}^{(m)}$, $\forall k$, and the channel vectors ${\bm h}_\text{MU}^\text{v}[k]$, $k = 1,\ldots,K$, can be recovered by using only a single subcarrier-independent measurement matrix. 
Observe, however, that the $\LBS$ and $\Lru$ available degrees of freedom at the transmitter and receiver are exploited.
Since \ac{OFDM} training pilots are forwarded through the \ac{MIMO} channel, the frequency-selective nature of the transmitted signal cannot be avoided. However, we can express the measurement matrix $\bm \Phi^{(m)}[k]$ as $\bm \Phi^{(m)}[k] = s^{(m)}[k]\bm \Phi^{(m)}$ if the OFDM symbols sent by the different \ac{MS}s are rotated versions of a baseline OFDM symbol. This can be written as $\bm s_u^{(m)}[k] = \bm q_u^{(m)} s^{(m)}[k]$, $\forall u$, with $\bm q_u^{(m)} \in \mathbb{C}^{\Lru \times 1}$ being a spatial modulation vector. With this choice, we can simultaneously exploit the degrees of freedom coming from having several RF chains at each \ac{MS} and have a frequency-independent measurement matrix, as shown hereafter.

Taking these considerations into account, the measurement matrix at the $m$-th training step can be written as
\begin{equation}
\bm \Phi^{(m)} = \left[\begin{array}{ccc}\bm q_1^{(m)T}\bm T_{\text{tr},1}^{(m)\Tr} &
\ldots & \bm q_U^{(m)T} \bm T_{\text{tr},U}^{(m)\Tr} \\ \end{array}\right] \otimes \bm F_\text{tr}^{(m)\He},
\label{equation:freq_ind_rx_frame}
\end{equation}
in which the pilot $s^{(m)}[k]$ is a subcarrier-dependent scalar whose effect can be eliminated at the receiver by simply multiplying the received vector by $(s^{(m)})^{-1}[k]$ or by $s^{(m)*}[k]$, where the latter would only apply for constellations with symbols having the same energy. Thereby, the subcarrier dependence of the matrix $\bm \Phi^{(m)}[k]$ in \eqref{equation:vect_rx_frame_channel} is avoided. It is important to remark that the different \ac{MS}s must use different precoders to transmit each training frame. The interpretation of such fact relies on the ability of the \ac{BS} to obtains enough information and discriminate the information of the channel for each \ac{MS}. In terms of sparse recovery, this means that users' using different precoders for a given training step translates into reduced correlation between the columns in $\bm \Phi^{(m)}$, thereby enabling support recovery for subsequent estimation of the channel gains and the channel matrices themselves.

If we consider the transmission of $M$ consecutive training frames, the received measurement vectors for each subcarrier can be written as an extension of \eqref{equation:freq_ind_rx_frame} as

\begin{eqnarray}
\underbrace{\left[\begin{array}{c} \bm y^{(1)\Tr}[k]  \\ \vdots \\ \bm y^{(M)\Tr}[k] \\ \end{array}\right]^\Tr}_{\bm y[k]} &\approx \underbrace{\left[\begin{array}{c} \bm \Phi^{(1)}  \\ \vdots \\ \bm \Phi^{(M)} \\ \end{array}\right]}_{\bm \Phi} \underbrace{\left( \bigoplus_{u=1}^{U}{\tilde{\bm \Psi}_u} \right)}_{\tilde{\bm \Psi}_\text{MU}} \tilde{\bm h}_\text{MU}[k] + \\ + & \underbrace{\left[\begin{array}{ccc} \bm n^{(1)\Tr}[k]  & \ldots & \bm n^{(M)\Tr}[k] \\ \end{array}\right]^\Tr}_{\bm n[k]}\nonumber,
\label{equation:full_signal_model}
\end{eqnarray}
where $\tilde{\bm \Psi}_u \in \mathbb{C}^{\Nt\Nru \times G_\text{BS} G_\text{MS}}$ is the dictionary matrix defined as $\tilde{\bm \Psi}_u \triangleq \left(\tilde{\bm A}_{\text{MS},u}^* \otimes \tilde{\bm A}_\text{BS}\right)$. The matrix $\tilde{\bm \Psi}_\text{MU} \in \mathbb{C}^{\Nt \sum_{u=1}^{U}{\Nru} \times U G_\text{BS} G_\text{MS}}$ is the multi-user dictionary. The $ML_\text{BS} \times 1$ vector $\bm n[k]$ is distributed according to ${\cal N}(\bm 0,\sigma^2 \bm C_\text{w})$, where $\bm C_\text{w}$ is the positive definite covariance matrix given by $\bm C_\text{w} = \blkdiag\left\{\bm F_\text{tr}^{(1)\He}\bm F_\text{tr}^{(1)},\ldots,\bm F_\text{tr}^{(M)\He}\bm F_\text{tr}^{(M)}\right\}$. Likewise, we also define the Cholesky factor of $\bm C_\text{w}$ as $\bm D_\text{w} \in \mathbb{C}^{M L_\text{r} \times M L_\text{r}}$, $\bm C_\text{w} = \bm D_\text{w}^\He \bm D_\text{w}$, which will be used in the proposed multi-user wideband channel estimation algorithm.

Therefore, the sparse vectors $\tilde{\bm h}_\text{MU}[k]$ can be recovered by solving the relaxed optimization problem

\begin{eqnarray}
\tilde{\bm h}_\text{MU}[k] = \underset{\bm h[k], k = 1,\ldots,K}{\arg\,\min\,} \sum_{k=1}^{K}{||\bm h[k]||_1}  \;\;\text{s.t.}\;\; \sum_{k=1}^{K}{L(\bm h[k])} \leq \epsilon\nonumber,
\end{eqnarray}
where 
\begin{equation}
L(\bm h[k]) = \left(\bm y[k] - \bm \Phi \tilde{\bm \Psi}_\text{MU} \bm h^\text{v}[k]\right)^\He \bm C_\text{w}^{-1}\left(\bm y[k] - \bm \Phi \tilde{\bm \Psi}_\text{MU} \bm h^\text{v}[k]\right)
\end{equation}
is the cost function associated to the \ac{LLF} of the received Gaussian signal \cite{Kay93}. The solution to this problem can be obtained by using either the \ac{SW-OMP} or the \ac{SS-SW-OMP+Th} algorithms in \cite{FreqChanEst17}, which have been shown to obtain estimation performance lying within the \ac{CRLB} when the AoA and AoD fall within the quantized spatial grid. We propose to use the \ac{SW-OMP} algorithm for illustration. In the next subsection, the \ac{CRLB} for the estimation of the channel matrices $\bm H_u[k]$, for all \ac{MS}s and subcarriers is provided. The channel estimation algorithm is provided in Algorithm \ref{alg:SWOMP}.

\begin{algorithm}
	\caption{Simultaneous Weighted Orthogonal Matching Pursuit (SW-OMP)}\label{alg:SWOMP}
	\begin{algorithmic}[1]
		\STATE  Initialize: 
		$\B{\Upsilon}_\text{w} \leftarrow \bm D_\text{w}^{-\He}\B{\Phi} \B{\Psi}$, $\bm y_\text{w}[k] \leftarrow \bm D_\text{w}^{-\He}\bm y[k]$, and $\bm r[k] \leftarrow \bm y_\text{w}[k]$, $k = 1,\ldots,K$ 
		\REPEAT 
		\STATE $\bm c[k] \leftarrow \B{\Upsilon}_\text{w}^\He \bm r[k]$, \quad $k = 1,\ldots,K$ 
		\STATE $p^\star \leftarrow \underset{p}{\arg\,\max}\,\sum_{k=1}^{K}{|[\bm c[k]]_p|}$ 
		\STATE  $\hat{\cal T} \leftarrow \hat{\cal T} \cup p^\star$ \hfill common support candidate
		\FORALL {$k$}
		\STATE $\bm x_{\hat{\cal T}}[k] \leftarrow \left(\left[\bm \Upsilon_\text{w}\right]_{:,\hat{\cal T}}\right)^\dag \bm y_\text{w}[k]$ \hfill support's subspace proj.
		\STATE $\bm r[k] \leftarrow \bm y_\text{w}[k]-\left[\bm \Upsilon_\text{w}\right]_{:,\hat{\cal T}} \bm x_{\hat{\cal T}}[k]$ \hfill update residual
		\ENDFOR
		\STATE	$\text{MSE} \leftarrow \frac{1}{KML_\text{r}}\sum_{k=1}^{K}{\bm r^\He[k] \bm r[k]}$
		\UNTIL{$\text{MSE}< \epsilon$}
	\end{algorithmic}
\end{algorithm}

\subsection{Cram\'{e}r-Rao Lower Bound}

This section is devoted to obtain the theoretical expression of the \ac{CRLB} for the estimation of the different subchannels for each \ac{MS}. We will focus on obtaining the bound assuming perfect estimation of the support of the multi-user channel vector. The motivation in doing so is to assess how well the estimator of the support performs, since the estimator of the channel gains derived in \cite{FreqChanEst17} is efficient if the support is correctly estimated. Assuming that the AoA and AoD have been found, the signal model in \eqref{equation:full_signal_model} reduces to
\begin{equation}
\bm y[k] = \overbrace{\bm \Phi \left(\bigoplus_{u=1}^{U}{\underbrace{\left(\bm A_{\text{MS}_u}^* \circ \bm {A}_{\text{BS}_u}\right)}_{\bm \Psi_u}}\right)}^{\bm \Upsilon} \bm \xi[k] + \bm n[k],
\end{equation}
with $\bm {A}_{\text{BS}_u}$ and $\bm A_{\text{MS}_u}$ the array matrices in \eqref{eq:equivCh} for the \ac{BS} and the $u$-th \ac{MS}. The $\sum_{u=1}^{U}{\Npu} \times 1$ vector $\bm \xi[k]$ contains the actual channel gains for the multi-user channel and is defined as 
\begin{equation}
\bm \xi[k] \triangleq \vect\left\{\left[\begin{array}{cccc} \diag\{\bm \Delta_1[k]\} & \ldots & \diag\{\bm \Delta_{U}[k]\} \\ \end{array}\right]\right\},
\end{equation}
where $\bm \Delta_u[k]$ are the $\Npu \times \Npu$ diagonal matrices in \eqref{eq:equivCh}. The Fisher Information Matrix for $\bm \xi[k]$ is given by \cite{Kay93},\cite{FreqChanEst17}
\begin{equation}
\bm I(\bm \xi[k]) = \frac{\bm \Upsilon^\He \bm C_\text{w}^{-1} \bm \Upsilon}{\sigma^2}.
\end{equation}
Thereby, the \ac{CRLB} matrix for the estimation of the multi-user channel matrix $\B{H}^\He[k]$ is given by the \textit{CRLB Theorem for Transformed Parameters} \cite{Kay93} as
\begin{equation}
\bm I^{-1}(\vect\{\B{H}^\He[k]\}) = \left(\bigoplus_{u=1}^{U}{\bm \Psi_u}\right) \bm I^{-1}(\bm \xi[k]) \left(\bigoplus_{u=1}^{U}{\bm \Psi_u^\He}\right).
\end{equation}
Finally, assuming that no prior information concerning the relation between the different channels $\bm H_u[k]$, $\forall k,u$ is available, the overall variance for the estimation of the $UK$ different channel matrices is denoted by $\gamma(\bm h_\text{total})$ and given by
\begin{equation}
\gamma(\bm h_\text{total}) = \sum_{k=1}^{K}{\trace\left\{\bm I^{-1}(\vect\{\bm H_u[k]\})\right\}},
\end{equation}
where $\bm h_\text{total} \in \mathbb{C}^{K N_\text{BS} \sum_{u=1}^UN_{\text{MS},u} \times 1}$ is given by 
\begin{equation}
\bm h_\text{total} = \vect\left\{\left[\begin{array}{cccc} \B{H}^\He[1] \ldots,\B{H}^\He[K] \\ \end{array}\right]\right\},
\end{equation}
such that the variance for the estimation of the different subchannels is added up owing to the lack of knowledge concerning possible correlation between the different channel matrices.

In the next subsection, we will introduce the different considerations to be made for a proper and fair comparison between uplink and downlink multi-user wideband channel estimation.

\subsection{Uplink vs. Downlink Multi-user Wideband Channel Estimation}

This subsection aims at providing a proper comparison between uplink and downlink \ac{mmWave} multi-user wideband channel estimation. The main problem is how to perform a comparison between both approaches comes from the different SNR measured at the \ac{MS}s and the \ac{BS}. Provided that the different signals transmitted by each \ac{MS} are uncorrelated, the SNR measured at the \ac{BS} will be the sum of the different SNRs concerning the individual signals transmitted by each \ac{MS}. This can be proven by considering a multi-user scenario with $U$ \ac{MS}. 

Let $\bm H_u^\He[k]$, $\bm T_{\text{tr},u}^{(m)}$, and $\bm s_u^{(m)}[k]$ denote the uplink channel matrix for the $u$-th user at subcarrier $k$, the training precoder used by the $u$-th user during the $m$-th training step, and the vector of training symbols sent by the $u$-th user during the $m$-th training step at $k$-th subcarrier, respectively. Let us also denote the received noise vector for the $m$-th training step and at the $k$-th subcarrier as $\bm n^{(m)}[k]$. Then, the received signal at the BS may be expressed as
\begin{equation}
	\bm y^{(m)}[k] = \underbrace{\sum_{u=1}^{U}\bm H_u^H[k] \bm T_{\text{tr},u}^{(m)}\bm s_u^{(m)}[k]}_{\bm x^{(m)}[k]} + \bm n^{(m)}[k].
\end{equation}
Next, let the receive SNR be defined as
\begin{equation}
	\text{SNR} = \frac{\trace\{\mathbb{E}\{\bm x^{(m)}[k] \bm x^{(m)H}[k]\}\}}{\trace\{\mathbb{E}\{\bm n^{(m)}[k]\bm n^{(m)H}[k]\}\}}.
	\label{equation:SNR_mu}
\end{equation}
Now, let us recall that the training symbols are assumed to be uncorrelated for the different users, i.e., $\mathbb{E}\{\bm s_u^{(m)}[k] \bm s_j^{(m)H}[k]\} = \bm 0$, for $u \neq j$. Accordingly, the cross-product terms in the numerator of \eqref{equation:SNR_mu} are zero-valued. Then, it turns out that the expression in \eqref{equation:SNR_mu} yields
\begin{equation}
	\text{SNR} = \sum_{u=1}^{U}\frac{P_u\trace\{\|\bm H_u^H[k]\bm F_{\text{tr},u}\|_F^2\}}{U N_\text{s} N_\text{r} \sigma^2},
\end{equation}
which is the summation of the SNR received by the different users, i.e., the summation of the SNR corresponding to the downlink for every user. 

Uncorrelation between the different signals can be claimed provided that the training precoders are pseudorandomly built with a network of phase-shifters having $N_\text{Q}$ quantization bits. Thereby, the comparison between UL and DL channel estimation will be made taking into account the SNR for a single user. This decision is made based on the fact that each \ac{MS} measures a SNR $U$ times smaller than that of the \ac{BS}. Yet, the \ac{MS}s separately estimate $K$ $\Npu$-sparse vectors each, while the \ac{BS} estimates $K$ $\sum_{u=1}^{U}{\Npu}$-sparse vectors, such that the proposed SNR criterion seems reasonable for comparison. In fact, the proposed criterion is the expected ratio between the SNR and the number of multipath components to be estimated, for either uplink or downlink scenarios.

\section{Numerical Results}
\label{sec:simulations}
In this section, we compute several numerical experiments to obtain further insight of the proposed methods. First, we present a performance comparison for the channel estimation in the uplink and the downlink in Sec. \ref{sec:channelEstPerf}. In Sec. \ref{sec:simulPCSI}, we evaluate the sum achievable rates obtained with digital and hybrid precoder/combiner designs considering perfect \ac{CSI}. Finally, Sec. \ref{sec:simulEstim} is devoted to experiments taking into account the available \ac{CSI}.

The setup considered for our experiments consist of a \ac{BS} with $\Nt=128$ antennas transmitting to $U=4$ \ac{MS}s deploying $\Nru=16$ antennas each, which individually allocate $\Nsu=2$ streams. The number of \ac{RF} chains are $\LBS=8$ and $\Lru=2$, while number of subcarriers considered is $K=32$. The channels are generated according to the $D$ delay model in \eqref{eq:channelModel}, with delay $D=8$ and $N_\text{p}=4$ paths.  The results are averaged by performing Monte Carlo simulations over $100$ channel realizations. We consider spatially withe noise for the uplink and the downlink, and its power is fixed according to  $\text{SNR}=P_\text{tx}/(U\sigma_n^2)$. The maximum number of iterations $\varepsilon$ is set to $200$ for Alg. \ref{alg:ProjGrad}, $100$ for HD-LSR and $50$ for the \ac{AM} methods.

\subsection{Channel Estimation Performance}
\label{sec:channelEstPerf}

This subsection is devoted to show simulation results concerning channel estimation performance for both \ac{UL} and \ac{DL} channel estimation. The NMSE expression in \eqref{eq:NMSE} will be used to compare the achievable performance when estimation is performed at the \ac{BS} and each of the $U$ \ac{MS}s. Besides the simulation parameters introduced at the beginning of this section, we take $\Gt = 2\Nt = 256$ and $\Gr = 2\Nr = 32$. The phase-shifters used in both the \ac{BS} and \ac{MS} are assumed to have $\NQ$ quantization bits, so that the entries of the training precoders and combiners $\bm F_\text{tr}^{(m)}$ and $\bm W_\text{tr}^{(m)}$, $m = 1,2,\ldots,M$ are drawn from a set ${\cal A} = \left\{0,\frac{2\pi}{2^\NQ}, \ldots,\frac{2\pi(2^{\NQ-1})}{2^\NQ}\right\}$. The number of quantization bits is set to $\NQ = 4$, and the number of training frames is set to $M = 100$.

\begin{figure}[ht]
		\centering
	\begin{subfigure}{0.85\linewidth}
		\includegraphics[width=\linewidth]{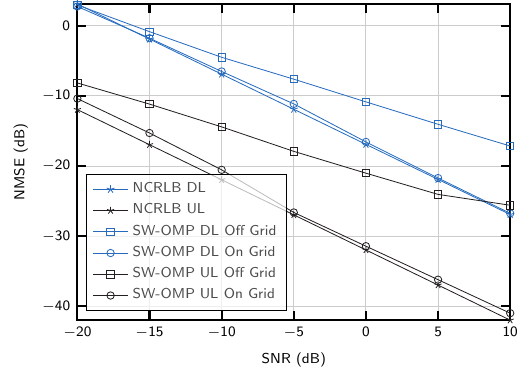}
		\caption{} 
		\label{fig:NMSE_vs_SNR}
	\end{subfigure}
	\begin{subfigure}{0.85\linewidth}
		\includegraphics[width=\linewidth]{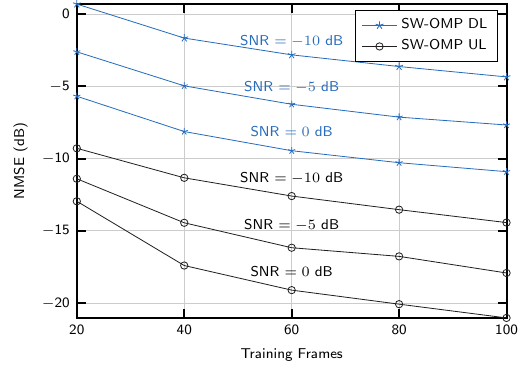}
		\caption{} 
		\label{fig:NMSE_vs_M}
	\end{subfigure}
	\caption{Estimation performance (NMSE) for both \ac{UL} and \ac{DL} multiuser channel estimation considering both on-grid and off-grid channel realizations: (a) NMSE versus SNR. (b) NMSE versus number of training frames $M$ for SNR ranging $\left\lbrace-10,-5,0\right\rbrace$ dB.}
	\label{fig:NMSE_Channel_Estimation}
\end{figure}

We show in Fig.~\ref{fig:NMSE_Channel_Estimation} the sample average NMSE versus both SNR and training overhead (number of training frames $M$). We first observe a large difference between channel estimation performance at the \ac{BS} and the \ac{MS}s. This is due to the higher amount of information the \ac{BS} can manage to estimate the channel. This comes from the larger dictionary used at the \ac{BS}, which makes it possible to perform a better compressive estimation. Furthermore, the \ac{BS} uses $\LBS = 8$ \ac{RF} chains for channel estimation at the, while each \ac{MS} is restricted to use $\Lru = 2$ \ac{RF} chains. This fact, jointly with the use of a larger dictionary, are the main reason why uplink channel estimation greatly outperforms its downlink counterpart. We also see that the gap between both approaches is reflected in the \ac{CRLB} for both cases. Both the estimation at the \ac{BS} and the \ac{MS}s exhibit performance lying very close to the \ac{CRLB}, although there is a small yet non-negligible gap between uplink channel estimation and the \ac{CRLB}, when the \ac{AoA}/\ac{AoD} fall within the quantized spatial grid. This gap can be further reduced by considering a larger number of subcarriers. This is because the sparsity level the \ac{BS} must estimate is approximately $U$ times larger (if the grid sizes are large enough), such that the estimation error of the channel gains is approximately $U$ larger as well. Therefore, according to the Law of Large Numbers, either a larger number of training frames or a larger number of subcarriers must be used to obtain better estimates of the noise variance. Thereby, the \ac{SW-OMP} algorithm can be halted more efficiently. This effect does not appear in the case of channel estimation in the \ac{DL}, since the sparsity level of each subchannel is lower. Nonetheless, it should be clear that, even with this small gap, channel estimation at the \ac{BS} greatly outperforms estimation at the \ac{MS}s. In the event of off-grid \ac{AoA}/\ac{AoD}, there is a considerable loss in performance for both \ac{UL} and \ac{DL} channel estimation, although the performance gap between them is approximately maintained. Further, for SNR ranging from $-10$ up to $0$ dB, the NMSE performance with off-grid parameters is lower than $-10$ dB, which is a clear indicator that the proposed estimation approach is suitable for multi-user \ac{UL} channel estimation with very-low SNR conditions.

\subsection{Performance Evaluation for Perfect CSI}
\label{sec:simulPCSI}

\begin{figure}[t]
		\centering
	\begin{subfigure}{0.85\linewidth}
		\includegraphics[width=\linewidth]{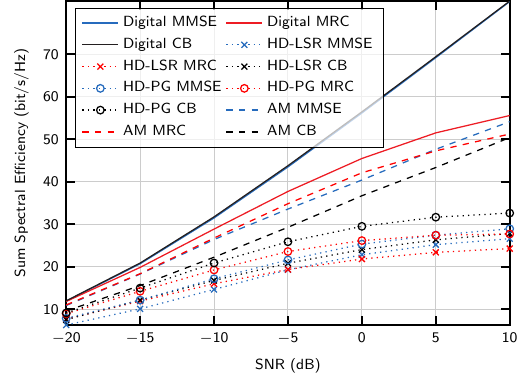}
		\caption{} 
		\label{fig:MMSEvsMRCvsZFHF}
	\end{subfigure}
	\begin{subfigure}{0.85\linewidth}
		\includegraphics[width=\linewidth]{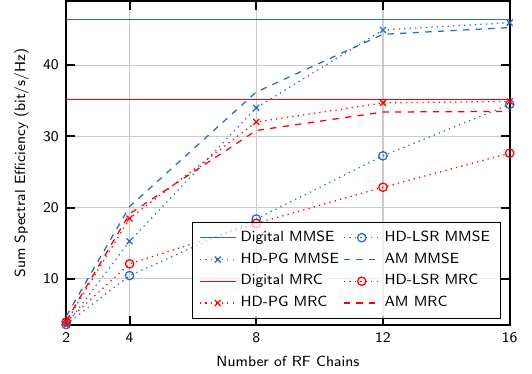}
		\caption{} 
		\label{fig:NRFChains}
	\end{subfigure}
	\caption{ (a) Achievable Sum-Rate vs. SNR for error-free channel estimation.  Digital \ac{MMSE}, Digital \ac{MRC}, and Digital \ac{CB} represent the digital benchmarks, HD-LSR and HD-PG are the factorizations using \cite[Least Squares relaxation]{RuMeGoHe16} and Alg. \ref{alg:ProjGrad}. Finally,  \ac{AM} \ac{MMSE}, \ac{AM} \ac{CB}, and \ac{AM} \ac{MRC} are the methods from Sec. \ref{sec:AM}. (b) Achievable Sum-Rate vs. N\# of RF Chains with $\LBS=2,4,8,12,16$. We consider $\Nt=64$, $\Lru=4$, and a fixed SNR of $0$dB.  }
	\label{fig:MMSEvsMRCvsZF}
\end{figure}

Fig. \ref{fig:MMSEvsMRCvsZFHF} compares the different digital designs in Sec. \ref{sec:digitalDesign} and the hybrid approaches in sections \ref{sec:DDFac} and \ref{sec:AM} for perfect \ac{CSI}. The performance of digital \ac{CB}, and \ac{MMSE} is very similar, whereas the gap for \ac{MRC} is increasing with the SNR.  Iterative methods in Sec. \ref{sec:AM} present better performance, specially for \ac{MMSE} combiner. Nevertheless, the performance of Alg. \ref{alg:ProjGrad} can be greatly improved using a smarter initialization, as stated in Sec. \ref{sec:simulEstim}. Fig. \ref{fig:NRFChains} shows that the performance of the digital solution can be achieved by the factorizations when the number of \ac{RF} chains is large enough, whereas the gap with respect to \ac{AM} is always greater than zero. Furthermore, less \ac{RF} chains are needed to obtain an accurate factorization for  \ac{MRC}, compared to that for \ac{MMSE}. 

\subsection{Performance Evaluation for Imperfect CSI}
\label{sec:simulEstim}
Now we evaluate the performance loss resulting from the imperfect \ac{CSI}. Let us start by studying the purely digital case to determine the robustness of the different combiner designs against the channel uncertainties. 
\begin{figure}[h]
	\centering
	\begin{subfigure}{0.85\linewidth}
		\includegraphics[width=\linewidth]{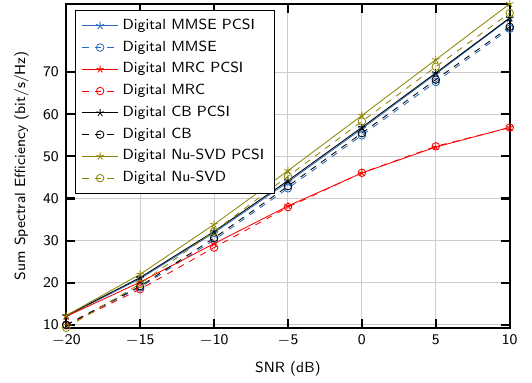}
		\caption{}
		\label{fig:MMSEvsMRCvsZFICSI}
	\end{subfigure}
	\begin{subfigure}{0.85\linewidth}
		\includegraphics[width=\linewidth]{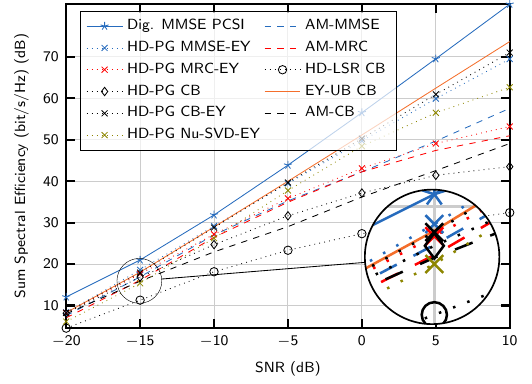}
		\caption{}
		\label{fig:HybridICSI}
	\end{subfigure}
	\caption{(a) Achievable Sum-Rate vs. SNR for perfect and imperfect \ac{CSI} using digital solutions with off-grid channel model. We include precoders and combiners jointly obtained with the algorithm \cite[Nu-SVD]{PaWoNg04} as benchmark. (b) Achievable Sum-Rate vs. SNR with hybrid precoders and combiners.   EY-UB CB represents the Eckart-Younglow rank approximation of CB. HD-PG and HD-LSR are the hybrid designs using Alg. \ref{alg:ProjGrad} and \cite[HD-LSR]{RuMeGoHe16}. MMSE-EY, CB-EY,  MRC-EY, Nu-SVD-EY correspond to Eckart-Young low rank approximations for the methods MMSE, CB,  maximum ratio combiner MRC, and Nu-SVD. AM are the alternating minimization methods.}
	\label{fig:ImperfectCSINp2}
\end{figure} 

	\begin{figure}[t]
		\centering
		\includegraphics[width=0.85\linewidth]{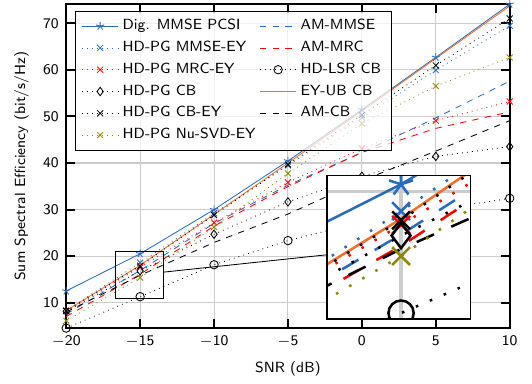}
	\caption{Achievable Sum-Rate vs. SNR with hybrid precoders and combiners. Compare with Fig. \ref{fig:HybridICSI}.}
	\label{fig:HybridICSINp4}
\end{figure}
As can be observed in Fig. \ref{fig:MMSEvsMRCvsZFICSI}, the gap introduced because of the estimation error is noticeable for \ac{MMSE}, \ac{CB} combiners. Contrarily, the robustness of \ac{MRC} leads to very similar performance for perfect and imperfect \ac{CSI}.

The gaps in Fig. \ref{fig:HybridICSI} come from the combination of several impairments that individually contribute to worsening the achievable performance. From the channel model in \eqref{eq:channelModelFreq}, it is clear that the rank of the resulting frequency-domain channel matrices increases with the number of paths $N_\text{p}$. Moreover, the multi-user uplink channel $\B{H}[k]=[\B{H}^\He_1[k]\ldots\B{H}^\He_U[k]]$ can be vectorized and expressed in terms of the array matrices employed by the \ac{BS} and MSs as $\vect\{\bm H[k]\} = \bigoplus_{u=1}^{U}\left(\bm A_{\text{MS},u}^* \circ \bm {A}_{\text{BS},u}\right) \vect\{\diag\{\bm \Delta_u[k]\}\}$. Then, if the AoA and AoD are i.i.d. and follow a continuous uniform distribution, and $\sum_{u=1}^{U}{\Npu} \leq \min(\Nt,\Nru)$,  the probability that transmit or receive steering vectors of the channels for different \ac{MS}s are linearly dependent is zero. Accordingly, $\vect\{\bm H[k]\}$ lies in a $\Nt \sum_{u=1}^{U}{\Nru}$-dimensional subspace spanned by $\sum_{u=1}^{U}{\Npu}$ non-orthogonal linearly independent vectors. Therefore, we can conclude that the rank of the multi-user channel grows linearly with the number of \ac{MS}s $U$ such that $\rank\{\bm H[k]\} = \min(N_\text{BS},\sum_{u=1}^{U}{\Npu})$. 

Observe that the lack of orthogonality of the steering vectors leads to inter-user interference.
Moreover, the proposed compressive approach aims at simultaneously estimating the different $UK$ subchannels without prior knowledge on the sparsity level. This brings about additional difficulty to accurately estimate the actual number of paths (i.e., the rank) of $\bm H[k]$. The estimation of a $UN_\text{p}$-sparse vector employing finite-resolution dictionaries leads to a larger grid quantization error compared to the single \ac{MS} case. Therefore, the estimation of the channel gains and, consequently, the noise variance, is more challenging. As a result, this creates a non-negligible performance gap between perfect and imperfect \ac{CSI}. 

Figure \ref{fig:HybridICSI} depicts the results obtained with the hybrid approaches.  Regarding the hybrid factorization problem, the main difficulty is that the analog \ac{UL} combiner (\ac{DL} precoder) must be designed to accommodate the different $K \Ns$ data streams. The number of available degrees of freedom of this analog precoder is $\LBS\ll K \Ns$, whereby it is very hard to design a low-rank hybrid precoder to simultaneously approximate $K U$ different rank-$\Ns$ matrices. Together with the stringent constraints of the \ac{RF} matrix, it makes hybrid factorization the system bottleneck.

After stacking the different combiners to perform the hybrid factorization, the rank of the overall combiner grows rapidly with \ac{MMSE} and Nu-SVD approaches. Since $N_\text{p} < \min(N_\text{BS},N_{\text{MS},u})$, $\forall u$, for a given \ac{MS}, the column spaces of digital \ac{MRC}s at the different subcarriers are equal to each other, up to a unitary rotation which is invariant in the Grassmannian manifold \cite{VeGoHe16}. Therefore, for \ac{MRC} the rank of the overall combiner does not sharply grow with the number of subcarriers $K$, yet it does with the number of users $U$ (see \eqref{eq:FMRC}). A similar effect is observed for \ac{CB} combiners in \eqref{eq:FZF}. Due to the limited rank of the inter-user interference matrix, the projection to the nullspace is approximately the identity matrix, i.e., $\B{U}_u[k]\B{U}_u[k]^\He\approx \mathbf{I}_{\Nt}$. By sharp contrast, provided that the inverse of the received covariance matrix in \eqref{eq:DLcomb} is different for each subcarrier, MMSE combiners destroy the relationship between column spaces of different subcarriers. Therefore, the rank of the overall \ac{UL} digital combiner grows both with $U$ and $K$, leading to the noticeable performance gap between the digital and the hybrid \ac{MMSE} combiners. This effect is even more accentuated in the case of Nu-SVD since the \ac{UL} precoders are not chosen as the right singular vectors of the channel estimate, as in the proposed digital designs.

The latter discussion indicates the suitability of \ac{MRC} for the low SNR regime expected at \ac{mmWave}. However, at low SNR, the diagonal elements of the matrix in \eqref{eq:DLcomb} are dominated by the noise term, whilst at high SNR the signal term  is dominant. Thereby, the lower the SNR, the less significant the alteration of the column space of the \ac{MMSE} combiner is.

Recall that equivalent overall \ac{DL} precoders and combiners have to be approximated by matrix products of ranks $\LBS=8$ and $\Lru=2$, respectively. Moreover, minimizing the Frobenius norm leads to near-optimal hybrid approximations \cite{AyRaAbPiHe14}. Therefore, by means of the Eckart-Young theorem \cite{EcYo36}, it is possible to set an upper bound for the performance of solutions based on factorization of digital designs. According to \cite{EcYo36}, for the digital precoder $\B{F}_\text{d}$ the error is upper bounded as  $\|\B{F}_\text{d}-\B{F}_\text{EY}\|_{\Frob}\leq\|\B{F}_\text{d}-\FRF\B{F}_\text{BB}\|_{\Frob}$ with the SVD $\B{F}_\text{d}=\B{U}_{\B{F}}\B{\Sigma}_{\B{F}}\B{V}_{\B{F}}^\He$, $\B{\Sigma}_\text{EY}$ containing the largest $\LBS$ values in the diagonal of $\B{\Sigma}_{\B{F}}$, and the optimal $\LBS$-rank approximation $\B{F}_\text{EY}=\B{U}_{\B{F}}\B{\Sigma}_\text{EY}\B{V}_{\B{F}}^\He$. Remarkably, by factorizing $\B{F}_\text{EY}$ instead of $\B{F}_\text{d}$, and initializing Alg. \ref{alg:ProjGrad} to $\FRF=\arg([\B{U}_{\B{F}}]_{1:\Nt,1:\LBS})$, the performance loss of the hybrid solution is mainly caused by the rank limitation and not by the factorization itself. Remember that the initialization is very important due to the local optimality of Alg. \ref{alg:ProjGrad}. The \ac{AM} approaches provide a good tradeoff among computational complexity and performance. It is very interesting for \ac{MRC} due to the good throughputs and computational efficiency.

Contrarily to the off-grid channel model, considering on-grid parameters does not result in the aforementioned quantization error, thereby simplifying channel estimation and hybrid precoder and combiner design. This follows from the fact that the probability of several paths sharing the same AoA and AoD in the multi-user channel is non-zero, being one if $U N_\text{p} \geq \min(G_\text{MS},G_\text{BS})$. Thereby, and jointly with on-grid delay paths, the rank of the multi-user channel does not necessarily increase with $U$ and $N_\text{p}$. Consequently, the difference in degrees of freedom between hybrid and digital solutions is reduced. Figure \ref{fig:HybridICSINp4} depicts the achievable rates for the case of on-grid \ac{AoA}/\ac{AoD} with hybrid precoders and combiners. Observe that both the estimation and factorization errors are negligible, and the overall performance is near-optimal under on-grid \ac{AoA}/\ac{AoD}.

\begin{table*}
	\renewcommand{\arraystretch}{1.3}
	\centering
	\caption{Performance of different approaches}
	\begin{tabular}{|c |c | c | c| c| c| c| c| c| c|}
		\cline{2-10}
		\multicolumn{1}{c|}{}& \multicolumn{3}{c|}{Digital} & \multicolumn{3}{c|}{HD-PG} & \multicolumn{3}{c|}{AM} \\
		 \cline{2-10} \multicolumn{1}{c|}{} &  MMSE & CB &  MRC & MMSE-EY &  CB-EY &  MRC-EY & MMSE & CB & MRC \\\hline
		Low-Mid SNR & Good & Good & Good &  Good & Good & Good & Good &  Poor & Good \\\hline
		Mid-High SNR & Good & Good & Mid & Good & Good & Mid & Mid &  Poor & Mid \\\hline
	\end{tabular}
	\label{tab:perf_summary}
\end{table*}   

\section{Conclusions}
\label{sec:conclusions}
This work tackles  channel estimation and the design of precoders and combiners in multi-user \ac{mmWave} \ac{MIMO} systems. The channel is estimated on the uplink, performed simultaneously for all the \ac{MS}s and subcarriers, and has been shown to greatly outperform its downlink counterpart for both on-grid and off-grid \ac{AoA}/\ac{AoD}. 
The precoders and combiners are designed according to different optimization criteria, presenting advantages depending on the SNR regime as shown in Table \ref{tab:perf_summary}. Specifically, although the \ac{MMSE} and \ac{CB} combiners perform better, the low complexity \ac{MRC} appears as an interesting alternative for low and mid SNRs. Simulation results exhibit the good performance of the proposed hybrid solutions, and the rank of the digital solutions is shown to be the performance-limiting factor. While the \ac{AM} approaches offers lower computational complexity, the digital factorization methods are more flexible and present better performance.     

\appendix
\subsection{Gradient Computation}
\label{ap:gradient}
To compute the gradient in \eqref{eq:gradStep}, we use the separation in \eqref{eq:distDecomp}, which yields 
\begin{align}
\frac{\partial d}{\partial \B{F}_\RF^*}&=\sum_{u=1}^{U}\sum_{k=1}^{K}\frac{\partial d_{k,l}}{\partial \B{F}_\RF^*},
\end{align}
allowing us to compute the derivative for each \ac{MS} and subcarrier, as follows
\begin{align*}
\frac{\partial d_{k,l}}{\partial \B{F}_\RF^*}&=-\frac{\sqrt{P_\text{tx}/(U\Nsu)}}{\|\B{A}_u[k]\|_{\Frob}}\frac{\partial\|\B{A}_u[k]\|_{\Frob}^2}{\partial \B{F}_\RF^*}\\
&=\frac{\sqrt{P_\text{tx}/(U\Nsu)}}{\|\B{A}_u[k]\|_{\Frob}}\big(\FRF(\FRFH\FRF)^{-1}\FRFH-\mathbf{I}_{\Nt}\big)\\&\times\B{F}^{u}_\MMSE[k]\B{F}^{u,\He}_\MMSE[k]\FRF(\FRFH\FRF)^{-1}.\nonumber
\end{align*}
Adding the individual contributions for each \ac{MS} and subcarrier we arrive at \eqref{eq:gradient}.

\subsection{Frobenius Norm for MSE}
\label{ap:Mseproof}
Let us introduce the matrix $\B{E}_u[k]=\B{F}^u_\MMSE[k]-\B{F}_\RF\FBBu[k]$ containing the difference between the unconstrained and the hybrid combiners. Now, we show that minimizing $\|\B{E}_u[k]\|_{\Frob}^2$ leads to a reduction in a lower bound for the \ac{MSE} in \eqref{eq:MSEULhybr}
\begin{align}
&\MSE^\UL_u[k]=\Nsu-2\real\{\trace\big((\B{F}^{u,\He}_\MMSE[k]-\B{E}_u[k]^\He)\B{R}^\UL_D[k]\big)\}\nonumber\\
&\quad+\trace\big((\B{F}^{u,\He}_\MMSE[k]-\B{E}_u^\He[k])\B{R}^{\UL}_I[k](\B{F}^u_\MMSE[k]-\B{E}_u[k])\big)\nonumber\\
&\quad=2\real\{\trace\big(\B{E}_u^\He[k]\B{R}^\UL_D[k]\big)\}+\trace\big(\B{E}_u^\He[k]\B{R}^{\UL}_I[k]\B{E}_u[k]\big)\nonumber\\
&\quad+\MMSE_u^\UL[k]-2\real\{\trace\big(\B{E}^\He_u[k]\B{R}^{\UL}_I[k]\B{F}^u_\MMSE[k]\big)\}\nonumber\\
&\quad=\MMSE_u^\UL[k]+\trace\big(\B{E}_u^\He[k]\B{R}^{\UL}_I[k]\B{E}_u[k]\big)\nonumber\\
&\quad\leq \MMSE_u^\UL[k]+\|\B{E}_u[k]\|^2_{\Frob}\|\B{R}^{\UL,1/2}_I[k]\|^2_{\Frob}.
\end{align}

\bibliographystyle{IEEEbib}  % Style BST file
\bibliography{refs}     % Bibliography file (usually '*.bib' ) 

%%%%%%%%%%%

\end{document}